\documentclass[12pt]{iopart}

\usepackage{subfigure}

\usepackage{graphicx}
\usepackage{mathrsfs}
\usepackage[british]{babel}
\usepackage{enumerate}
\usepackage{enumitem}
\usepackage{bbm}

\usepackage[sort&compress,numbers]{natbib}
\usepackage{doi}
\usepackage{hyperref}

\usepackage{comment} 
\expandafter\let\csname equation*\endcsname\relax 
\expandafter\let\csname endequation*\endcsname\relax 

\usepackage{amsbsy,amssymb,amsmath,bm}
\usepackage{graphicx,color,epsfig,rotate}

\catcode`@=11 
\renewcommand\tableofcontents{%
  \section*{\contentsname}%
  \@starttoc{toc}%
}
\catcode`@=12

\def\XXint#1#2#3{{\setbox0=\hbox{$#1{#2#3}{\int}$}
     \vcenter{\hbox{$#2#3$}}\kern-.5\wd0}}

\def\intc{c}
\def\pot{\alpha}

\def\qq{{\rm q}}
\def\jj{{\rm j}}

\def\barj{{j_0}}

\newcommand{\Ket}[1]{\left|#1  \right>}

\renewcommand{\imath}{i}

\newcommand{\beq}{\begin{equation}}
\newcommand{\eeq}{\end{equation}}

\usepackage{xcolor}

\newcommand{\dd}{d}

\usepackage{multirow}
\newcommand{\be}{\begin{equation}}
\newcommand{\ee}{\end{equation}}

\usepackage{etoolbox}

\makeatletter
\def\@mkboth#1#2{}
\newlength\appendixwidth
\preto\appendix{\addtocontents{toc}{\protect\patchl@section}}
\newcommand{\patchl@section}{%
  \settowidth{\appendixwidth}{\textbf{Appendix }}%
  \addtolength{\appendixwidth}{1.5em}%
  \patchcmd{\l@section}{1.5em}{\appendixwidth}{}{\ddt}%
}
\makeatother

\begin{document}

\title{Hydrodynamics of weak integrability breaking}

\author{Alvise Bastianello}
\address{Department of Physics and Institute for Advanced Study,
Technical University of Munich, 85748 Garching, Germany}
\address{and Munich Center for Quantum Science and Technology (MCQST), Schellingstr. 4, D-80799  M\"unchen, Germany}

\author{Andrea De Luca}
\address{Laboratoire de Physique Th\'eorique et Mod\'elisation (UMR 8089), CY Cergy Paris Universit\'e, CNRS, F-95302 Cergy-Pontoise, France}

\author{Romain Vasseur}
\address{Department of Physics, University of Massachusetts, Amherst, MA 01003, USA }


\begin{abstract} 
 We review recent progress in understanding nearly integrable models within the framework of generalized hydrodynamics (GHD).
 Integrable systems have infinitely many conserved quantities and stable quasiparticle excitations: when integrability is broken, only a few residual conserved quantities survive, eventually leading to thermalization, chaotic dynamics and conventional hydrodynamics. In this review, we summarize recent efforts to take into account small integrability breaking terms, and describe the transition from GHD to standard hydrodynamics. We discuss the current state of the art, with emphasis on weakly inhomogeneous potentials, generalized Boltzmann equations and collision integrals, as well as bound-state recombination effects. We also identify important open questions for future works. 
 
\end{abstract}

\maketitle

\newpage

\tableofcontents

\section{Introduction}

In one dimension, many paradigmatic models of condensed matter physics are integrable, including the Hubbard, Heisenberg, and Lieb-Liniger models. Although such systems are ``exactly-solvable'' in some sense, their non-equilibrium and dynamical finite-temperature properties have only started to be uncovered in recent years \cite{Calabrese_2016} at the price of huge efforts.
Furthermore, the apparent fragility of integrability against inhomogeneities and external perturbations, which are ubiquitous in real-life experiments, can potentially be a bottleneck to its quantitative experimental applications. 
This quest endured until the recent breakthrough of ``Generalized Hydrodynamics'' (GHD)~\cite{GHD1, GHD2}, which takes into account the presence of infinitely many conservation laws and of stable  quasiparticle excitations characteristic of integrable systems. GHD allowed the elegant methods of integrability to describe quantitatively cold atom experiments in highly excited and far from equilibrium setups \cite{PhysRevLett.122.090601,malvania2020generalized}, sparkling interest in this new approach.

GHD has led to a quantitative understanding of non-equilibrium transport setups~\cite{GHD1, GHD2, SciPostPhys.2.2.014,PhysRevB.96.115124, 10.21468/SciPostPhys.6.6.070, DOYON2018570, Doyon_2017, PhysRevLett.120.176801, PhysRevB.97.081111, PhysRevLett.119.220604,PhysRevLett.119.195301, BertiniPiroli_2018,PhysRevB.98.075421, PhysRevLett.120.045301, PhysRevLett.123.130602,10.21468/SciPostPhys.8.3.041,PhysRevB.100.035108, PhysRevB.96.220302, 10.21468/SciPostPhys.8.1.007, DoyonMyers, DoyonToda, Bulchandani_2019Toda, Cao_2019Toda, PhysRevLett.124.140603,PhysRevX.10.011054,PhysRevLett.125.070602,10.21468/SciPostPhys.8.2.016,PhysRevE.102.042128,PhysRevB.101.035121,PhysRevB.103.035130}, entanglement \cite{Alba7947,Bertini_2018,10.21468/SciPostPhys.7.1.005} and correlation spreading \cite{10.21468/SciPostPhys.5.5.054,10.21468/SciPostPhysCore.3.2.016}, as well as analytical expressions for linear response quantities such as Drude weights~\cite{PhysRevLett.119.020602,BBH, GHDDrudeBoseGas, PhysRevB.96.081118, PhysRevB.97.081111} and diffusion constants~\cite{PhysRevLett.121.160603, PhysRevB.98.220303, 10.21468/SciPostPhys.6.4.049,gv_superdiffusion,10.21468/SciPostPhys.9.5.075,2019arXiv191201551D}. GHD also revealed the existence of anomalous transport regimes in strongly interacting spin chains~\cite{PhysRevLett.106.220601,lzp,idmp,gv_superdiffusion,PhysRevLett.123.186601,gvw, dupont_moore,vir2019, 2019arXiv190905263A,PhysRevLett.122.210602,PhysRevB.102.115121,2020arXiv200908425I,dmki,PhysRevLett.125.070601}, which  motivated recent experiments~\cite{Jepsen2020,PhysRevLett.113.147205,2020arXiv200913535S}.

However, realistic systems are only approximately integrable, and will always include small perturbations breaking integrability. Integrability-breaking perturbations can also be tuned experimentally~\cite{PhysRevX.8.021030}, providing an interesting knob to interpolate between integrable and chaotic dynamics.
At short times, such systems should be well described by GHD, and we expect transport properties to be generically {\em ballistic}. 
In contrast, non-integrable systems only have a few conservation laws (typically energy, particle number, and/or momentum), and do not support stable quasiparticles at high temperature.
At long enough times, the dynamics should therefore become chaotic and a crossover to conventional (primarily {\em diffusive} unless other symmetries are present) hydrodynamics should occur. There are many interesting and fundamental questions associated with breaking integrability, in particular related to level statistics, quenches and other probes of quantum chaos (see {\it e.g.}~\cite{doi:10.1080/00018732.2016.1198134,PhysRevX.5.041043,2017arXiv170602031S,2020arXiv201207849L} for recent progress), and in this review we will focus on this question from the perspective of the theory of hydrodynamics. 

Qualitatively, the effects of weak integrability breaking are rather clear, and have been studied numerically for many years (see {\it e.g.}~\cite{PhysRevB.53.983, PhysRevB.76.245108,PhysRevLett.110.070602,PhysRevB.91.115130,PhysRevB.90.094417, PhysRevB.93.205121} among many others). The perturbation can either act at first order in perturbation theory through forces due to inhomogeneous potentials, or at second order in perturbation theory leading to the decay of most charges that were conserved in the integrable limit. Integrability-breaking perturbations also endow quasiparticles with a finite lifetime, so quasiparticle excitations can now scatter diffractively, decay and/or backscatter depending on the precise nature of the perturbation. However, describing those processes quantitatively for strongly interacting systems presents a daunting challenge that has only started to be addressed in recent years. Note that integrability breaking has been studied for decades in weakly-interacting systems, Fermi and Luttinger liquids, and is relatively well understood by now (see {\it e.g.}~\cite{landau1981course,kamenev2011field,PhysRevLett.115.180601,PhysRevB.94.245117,PhysRevLett.85.1092,PhysRevLett.103.216602,PhysRevB.83.035115,PhysRevB.88.115126,PhysRevLett.103.096402,PhysRevB.97.045105,Bulchandani12713,
PhysRevE.86.031122,PhysRevE.88.012108,F_rst_2013}). In this review, we will focus on interacting settings, where GHD has led to important progress in the past couple of years.

A natural route to address weakly decaying quasiparticles -- inspired from the theory of weakly interacting classical and quantum gases -- is the Boltzmann equation \cite{10.1007/BFb0071883,PhysRevE.88.012108,PhysRevE.86.031122,Spohn2006,PhysRevB.88.115144,PhysRevA.89.053608,PhysRevLett.115.180601,PhysRevB.94.245117, Biella2019}. Since the perturbation's effects are smooth on thermodynamic properties (like the equation of states), but singular on dynamical properties at long times, to leading order one can assume that the system is always locally in a generalized equilibrium state of the integrable system. The integrability breaking perturbation then leads to force terms, and to collision integrals which rely on the matrix elements of generic local operators (``form factors'') between eigenstates of the unperturbed (integrable) system. Determining these matrix elements is a challenging task despite some recent progress \cite{De_Nardis_2018,CortesCubero2019,10.21468/SciPostPhys.8.1.004, G_hmann_2017, Kitanine_2011}, and physically-motivated approximations are likely to be needed.

The rest of this review is organized as follows. In the remainder of the introduction section, we provide a short summary of basic integrability concepts and notation, to help the reader to quickly navigate this review.
In section~\ref{sec_force_first_order}, we first describe first-order effects of integrability breaking (generalized forces) due to inhomogeneous potentials. As an example, we discuss the case of relaxation of the 1d Bose gas in a trap, and show how thermalization can occur by including diffusive corrections to GHD combined with the presence of the trap. In section~\ref{sectionBoltzmann}, we then move on to second-order effects and introduce a generalized Boltzmann equation framework that consists of adding collision integral terms to GHD equations. We discuss how this generically leads to diffusive hydrodynamics and compute the resulting diffusion constants for the residual charges perturbatively. We illustrate this general approach on the cases of atom losses and smoothly varying noise, where the ``collision integrals'' and relevant matrix elements can be evaluated exactly, before briefly discussing possible approximations.  
Finally, in section~\ref{sec_hidd_nonadiab}, we discuss the case of integrable systems whose set of conserved charges varies discontinuously with a tuning parameter. Qualitatively, this implies the simultaneous closure, or opening, of infinitely many gaps among the conserved charges. Within a collision-integral interpretation, this gap closure can be seen as a resonance in the Fermi-Golden rule and the excitations are quickly rearranged in the allowed phase-space, leading to bound state recombination. We conclude by highlighting important questions for future works. 
Finally, we discuss stable numerical methods to solve hydrodynamic equations in a brief appendix.

\subsection{List of notations}

For the sake of clarity and to offer the reader a quick grasp on the notation, we provide a short summary of the rudiments of the thermodynamic description of integrable models, together with a list of notations here used. For a more exhaustive introduction, the reader can refer to the literature, for example Refs. \cite{takahashi2005thermodynamics,franchini2017introduction}.
Homogeneous integrable models are characterized by an extensive set of (quasi-)local charges $\{Q_j\}_j$
\be
Q_j=\int\dd x\,  {\rm q}_j(x)\, ,
\ee
which are linearly independent, conserved by the dynamics $[H,Q_j]=0$ and are in involution $[Q_i,Q_j]=0$. For the sake of simplicity, we imagine a continuous model, but the same concepts hold for systems on a lattice as well.
The operator ${\rm q}_j(x)$ is called the charge density and it usually has a compact support for local charges. In addition to local charges, several integrable models exhibit  quasi-local charges as well, in which case ${\rm q}_j(x)$ has an exponentially-decaying support \cite{Ilievski_2016}.
Since the charge is a conserved quantity, its density obeys the continuity equation
\be\label{eq_chj_continuity}
\partial_t \qq_j(x)+\partial_x \jj_j(x)=0\, ,
\ee
with $\jj_j(x)$ being the associated current density.
The presence of conservation laws prevents the system from thermalizing in the usual sense, as the system will instead relax to a so-called Generalized Gibbs Ensemble~\cite{PhysRevLett.98.050405,Vidmar_2016, Doyon2017} described by a density matrix $\hat{\rho}$ 
\be
\hat{\rho}\propto e^{-\sum_j\beta_j Q_j}
\ee

Another striking effect of the infinitely many conservation laws is the presence of stable particle-like excitations, which undergo completely elastic scatterings despite the strongly interacting nature of the system. In the thermodynamic limit, GGEs are uniquely identified by the phase-space densities of these excitations \cite{PhysRevLett.115.157201, PhysRevLett.110.257203} within the framework of the Thermodynamic Bethe Ansatz (TBA) \cite{takahashi2005thermodynamics}.
These counting functions are called root densities and, following the standard conventions, are denoted as $\{\rho_j(\lambda)\}_j$, where $\lambda$ is the rapidity of the excitation, while the index $j$ runs over the excitations species, called also strings. On a GGE, the expectation value of a charge density is expressed as
\be
q_j(x) \equiv \langle {\rm q}_j(x)\rangle_\text{GGE}=\sum_{i}\int \dd\lambda \, h_{j,i} (\lambda)\rho_i(\lambda)\,,
\ee
where $h_{j,i}(\lambda)$ is the charge eigenvalue of the charge ${\rm q}_j(x)$ relative to the string $i$. The charge eigenvalue depends on the details of the model and of the charge, but it is state-independent, which in turn is described by the root density.

It is useful to introduce a separated notation for the charge eigenvalues of the Hamiltonian and the momentum, which are respectively called $\epsilon_j(\lambda)$ and $p_j(\lambda)$.

Despite the elasticity of the scattering among the quasiparticles, the latter are strongly interacting and they get a phase shift whenever they scatter. This affects the finite-volume quantization and hence the allowed phase-space, which becomes state dependent.
This naturally leads to introduce the so called total root density $\rho_j^t(\lambda)$ which accounts for the allowed phase space density for the $j^\text{th}$-string at rapidity $\lambda$
\be
\rho_j^t(\lambda)=\frac{\partial_\lambda p_j(\lambda)}{2\pi}-\sum_{j'}\int \frac{\dd\lambda'}{2\pi} \, \partial_\lambda \Theta_{j,j'}(\lambda-\lambda')\rho_{j'}(\lambda')\, .
\ee
Above, $\Theta_{j,j'}(\lambda)$ is the scattering phase, which captures the interactions and is model dependent. Notice that, in the absence of interactions $\Theta=0$, $\rho_j^t$ becomes the phase-space density of free particles in a box. 
For the sake of notation, it is useful to define $\varphi_{j,j'}(\lambda)= \partial_\lambda\Theta_{j,j'}(\lambda)$ and the dressing operation $\tau_j(\lambda)\to \tau_j^\text{dr}(\lambda)$
\be\label{def_dr}
\tau_j^\text{dr}(\lambda)=\tau_j(\lambda)-\sum_{j'}\int \frac{\dd \lambda'}{2\pi}\varphi_{j,j'}(\lambda-\lambda') n_{j'}(\lambda') \tau_{j'}^\text{dr}(\lambda')\, ,
\ee
where $0\le n_j(\lambda)\le 1$ is the so called filling fraction
\be\label{eq_fillfrac}
n_j(\lambda)=\frac{\rho_j(\lambda)}{\rho_j^t(\lambda)}\, .
\ee
The dressing operation is ofter written more compactly in the operator notation $\tau^\text{dr} = (1 + \frac{\boldsymbol{\varphi} \bf{n}}{2\pi})^{-1} \tau$, where the action of a matrix $A$ is defined as
\begin{equation}
\label{eq:opnot}
 \boldsymbol{A} f = \sum_{j'} \int d\lambda' \boldsymbol{A}_{j,j'}(\lambda, \lambda') f(\lambda') 
\end{equation}
and $\boldsymbol{\varphi}_{j'j'}(\lambda, \lambda') = \varphi_{j,j'}(\lambda - \lambda')$, $\boldsymbol{n}_{j,j'}(\lambda, \lambda') = \delta(\lambda - \lambda') \delta_{jj'} n_j(\lambda)$.

The thermodynamics of the system, the GGEs and the whole generalized hydrodynamics, are built upon these quantities.

\section{Weakly inhomogeneous couplings}
\label{sec_inh_couplings}

Integrability is fragile against inhomogeneities in the Hamiltonian, which eventually cause the system to be non-integrable when seen as a whole. However, in the case of very weak inhomogeneities, the local dynamics is still approximately integrable and one can use the hydrodynamic approach.

In this section, we will review the advances in including weak spatial inhomogeneities and time dependence in the dynamics.
More precisely, let $H(\alpha)\equiv\int \dd x\, {\rm h}(x,\alpha)$ be the Hamiltonians of a family of integrable models, parametrized through a certain parameter $\alpha$. In the same notation, let $Q_j(\alpha)\equiv\int \dd x\, {\rm q}_j(x,\alpha)$ be the conserved charges where the parametric dependence on $\alpha$ has been made explicit.
The coupling can be promoted to be weakly space-time dependent $\alpha\to \alpha(t,x)$, resulting in an inhomogeneous and time-dependent dynamics for the whole system
\be
\label{eq:halpha}
H=\int \dd x\, {\rm h}(x,\alpha(t,x))\, .
\ee
The state describing the system is supposed to be inhomogeneous as well, but locally described by the GGE based on the Hamiltonian $H(\alpha(t,x))$.
We stress that with $H(\alpha(t,x))$ we denote the family of homogeneous Hamiltonian $H(\alpha)$ we previously defined with the replacement $\alpha\to \alpha(t,x)$. Therefore, $H(\alpha(t,x))$ defines an homogeneous integrable model such that nearby $(t,x)$ it describes the same dynamics as the Hamiltonian of interest \eqref{eq:halpha} (which in turn is non-integrable due to the inhomogeneity).
So far, we considered generic inhomogeneities, but it is useful to further divide them in two distinct classes
\begin{enumerate}
\item \label{Case1} Inhomogeneities such that the local Hamiltonians commute $[H(\alpha(t,x)),H(\alpha(t',x'))]=0$, as well as all the (quasi-)local conserved charges. In this case, one can fix a reference point, e.g. $(t,x)=(0,0)$, then the vanishing commutator, together with the completeness of the set of charges, tells us that the Hamiltonian $H(\alpha(t,x))$ must be a linear combination of conserved charges of the model in the origin.
\be\label{eq_Dinh}
H(\alpha(t,x))= \sum_j c_j(\alpha(t,x)) \int \dd y  \; {\rm q}_j(y,\alpha(0,0))\, ,
\ee
with $c_j(\alpha(t,x))$ some coefficients. In other words, the integrable model which locally describes the dynamics does not change, since the set of local conserved quantities remains the same, but the generator of the time evolution rotates in the space of the conserved charges.
As a direct consequence, the scattering matrix and the excitations' content, i.e. the strings, do not change. This is important in the perspective of Section \ref{sec_hidd_nonadiab} when we discuss bound state recombination: this effect cannot be attained with inhomogeneities in the form Eq. \eqref{eq_Dinh}.
In this setup, the requirement of smooth inhomogeneities is simply translated into $c_j$ being smooth functions of their argument.

\item \label{Case2} The Hamiltonians, as well as the charges, do not commute any longer $[H(\alpha(t,x)),H(\alpha(t',x'))] \neq 0$. 
In this case, the integrable model which locally describes the dynamics of the system changes in space and time. As a consequence, the scattering data and in principle the excitations' content can be a function of space and time: extra care should be taken in the notion of smooth inhomogeneities.
Precisely, we say the dependence on $\alpha$ is smooth if the whole set of charge densities ${\rm q}_j(x,\alpha)$ is smooth in $\alpha$.
Notice that integrable models are very sensitive to the value of the interactions and inhomogeneities that look smooth at the level of the Hamiltonian density could have drastic changes on the other charges. The case of these ``hidden inhomogeneities'' requires a different analysis and will be considered in Section \ref{sec_hidd_nonadiab}.
\end{enumerate}

This section is structured as follows.
In Section \ref{sec_force_first_order} we discuss in details the  hydrodynamic equations up to the first order in the derivative expansion of the inhomogeneity. These results, in the two cases described above, have been obtained in Ref. \cite{SciPostPhys.2.2.014} and Ref. \cite{PhysRevLett.123.130602} respectively.
Then, in Section \ref{sec_firstorder_lackth} we overview its applications and discuss the consequences on integrability-breaking and lack of thermalization within this approximation \cite{10.21468/SciPostPhys.6.6.070,PhysRevLett.120.164101}.
Section \ref{sec_force_second_order} discusses the late-time thermalization due to diffusive corrections \cite{PhysRevLett.125.240604}.
\ref{sec_num} contains a short overview of efficient and stable numerical methods that can be used to solve the hydrodynamic equations.

\subsection{Force fields from first order effects}
\label{sec_force_first_order}

In this section we review the general form of GHD in the presence of inhomogeneities in the dynamics, causing the appearance of force terms in the GHD equations \cite{SciPostPhys.2.2.014,PhysRevLett.123.130602}.
On general grounds, hydrodynamics deals with slow and long wavelength modes in the system: first, one invokes a local density approximation by claiming the system has locally relaxed to a steady state. Since the dynamics is locally integrable, the system attains local relaxation to a GGE which is conveniently described by a local root density (``quasiparticle distribution''). Physically, the inhomogeneous root density is nothing else than the phase space density of the excitations.
In a second step, the slow hydrodynamic evolution is considered by writing the proper hydrodynamic equations for the root density. The leitmotiv of this approach is the assumption of a separation of space-time scales: as a consequence, the hydrodynamic equations can be organized in a gradient expansion in the inhomogeneity of the state and of the interactions.
Within this section, we focus solely on the first order, also called as Eulerian GHD: diffusive corrections beyond this approximation are extremely important for the late time dynamics and thermalization, and will be discussed in Section \ref{sec_force_second_order}.

Before embarking into the derivation of the GHD equations, we discuss their general form and their physical content. In this section, we do not focus on a particular system and we present results valid in general.
At the Eulerean scale, the GHD equation is a simple continuity equation for particles moving with velocity $v^\text{eff}$ and feeling a force $F^\text{eff}$ (we leave the spatial and rapidity dependence implicit for the sake of notation). 
\be\label{eq_ghd_1ord}
\partial_t \rho+\partial_x (v^\text{eff}\rho)+\partial_\lambda (F^\text{eff}\rho)=0\, .
\ee
For simplicity, we consider the case where a single particle species is present, the generalization to several species is trivially obtained adding the string dependence to the root density, effective velocity and effective force.
In the case of a homogeneous and time independent Hamiltonian the force term is absent $F^\text{eff}=0$: inhomogeneous potentials or interactions induce a non trivial force term proportional to the space-time derivatives of the inhomogeneity. Altogether, Eq. \eqref{eq_ghd_1ord} is a first order equation in space-time derivatives.
Equivalently, the GHD equation can be rewritten in terms of the filling fraction as 
\begin{equation}
\label{eq_ghd_1ord_filling}
\partial_t n+ v^\text{eff}\partial_x n+F^\text{eff}\partial_\lambda n=0    \, .
\end{equation}
In the absence of forces, this form of the GHD equations has already been considered in the seminal papers \cite{GHD1, GHD2} and can be obtained by applying Eq. \eqref{eq_ghd_1ord} to the definition of the filling fraction \eqref{eq_fillfrac}. In the presence of forces, Eq. \eqref{eq_ghd_1ord_filling} has been obtained from Eq. \eqref{eq_ghd_1ord} in Refs. \cite{SciPostPhys.2.2.014,PhysRevLett.123.130602}.
The form Eq. \eqref{eq_ghd_1ord_filling} is more convenient for numerical purposes, since the method of characteristics provides a stable algorithm for its solution (see \ref{sec_num}).
The effective velocity is defined as
\be\label{eq_veff}
v^\text{eff}=(\partial_\lambda \epsilon)^\text{dr}/(\partial_\lambda p)^\text{dr},
\ee
and is the group velocity of the excitations renormalized by the interactions. 
This ansatz for the effective velocity is the central result of the seminal papers \cite{GHD1, GHD2}: in Ref. \cite{GHD1} the form \eqref{eq_veff} has been proposed on the basis of a heuristic kinetic picture, while Ref.  \cite{GHD2} offers a derivation in relativistic-invariant systems. Later on, the expression \eqref{eq_veff} has been rigorously proven in Refs. \cite{PhysRevLett.125.070602,PhysRevX.10.011054}.
The focus of this review is on the effect of explicitly integrability breaking terms, which lead to the effective force $F^\text{eff}$: we leave a more extensive discussion of the effective velocity to the reviews \cite{denardis2021correlation,alba2021generalizedhydrodynamic,borsi2021current} in the same volume.
The effective velocity acquires a space and time dependence due to two effects, one implicit and the other explicit. The implicit dependence is due to the dressing \eqref{def_dr}, which makes the effective velocity state-dependent. The explicit dependence is due to the fact that the dispersion law of the (locally-)integrable model in general depends on the coupling $\alpha$, therefore it develops a parametric inhomogeneity $\alpha\to \alpha(t,x)$.
A word of caution should be given comparing the current convention with that of Ref. \cite{SciPostPhys.2.2.014}, where inhomogeneities in the form (\ref{Case1}) were considered. In this case, the inhomogeneous Hamiltonian can be written as $H=\int\dd x \{{\rm h}(x)+\alpha(t,x){\rm q}(x)\}$, for some charge density ${\rm q}$ (or linear combinations thereof). Hence, in this case the local energy eigenvalue is $\epsilon^{(\alpha(t,x))}(\lambda)=\epsilon^{(0)}(\lambda)+\alpha(t,x) h(\lambda)$, where $h(\lambda)$ is the eigenvalue of the charge density ${\rm q(x)}$ and $\epsilon^{(0)}$ is the dispersion law in the absence of the inhomogeneous field $\alpha=0$. In Ref. \cite{SciPostPhys.2.2.014} the authors kept the expression for $\epsilon^{(\alpha)}$ explicit, while in this section we rather follow the convention of Ref. \cite{PhysRevLett.123.130602} and $\epsilon$ is meant to be the dispersion law of the full local Hamiltonian.

The effective force receives two contributions, one coupled to spatial inhomogeneities and the other to temporal changes
\be\label{eq_F}
F^\text{eff}=[\partial_t\alpha f^\text{dr}+\partial_x\alpha \Lambda^\text{dr}]/(\partial_\lambda p)^\text{dr}\,,
\ee
where
\begin{eqnarray}\label{eq_f1}
f(\lambda)&=&-\partial_\alpha p(\lambda)+\int \frac{\dd\lambda'}{2\pi}\partial_\alpha \Theta(\lambda-\lambda')(\partial_{\lambda'}p)^\text{dr}n(\lambda'),\\
\label{eq_f2}
\Lambda(\lambda)&=&-\partial_\alpha \epsilon(\lambda)+\int \frac{\dd\lambda'}{2\pi}\partial_\alpha \Theta(\lambda-\lambda')(\partial_{\lambda'}\epsilon)^\text{dr}n(\lambda').
\end{eqnarray}
Notice that $F^\text{eff}$ is explicitly linear in the space and time derivatives, as we have already anticipated.
In the expressions for $f$ and $\Lambda$, one can recognize two physically distinct contributions. The first terms, i.e. the $\alpha-$derivatives of the momentum and energy eigenvalues, can be understood as single particle effects, i.e. the inhomogeneity causes a space-time dependent dispersion law with the consequence of accelerating the particle. This effect is present also in free models and the role of the interaction is entirely captured by the dressing.
On the other hand, the second contributions described by the integrals in Eqs. (\ref{eq_f1}-\ref{eq_f2}) are intrinsically many-body effects. Indeed, $\partial_\alpha\Theta$ is non-vanishing only if the inhomogeneity modifies the scattering data, i.e. the interactions among the particles. Furthermore, these terms are proportional to the filling fraction $n(\lambda)$ and vanishes in the low-density limit.
It is worth mentioning that the force terms greatly simplify if one considers inhomogeneities in the form (\ref{Case1}): in this case, the inhomogeneity does not modify the scattering data $\partial_\alpha\Theta=0$, neither the momentum $\partial_\alpha p=0$, but only affects the local energy eigenvalue $\partial_\alpha \epsilon\ne 0$.
As it is self-evident from Eqs. (\ref{eq_ghd_1ord}-\ref{eq_F}), the GHD equations contain only first order derivatives in time and space, therefore they are invariant under a global rescaling of space and time $(t,x)\to (A t,A x)$. This symmetry is of course not exact in the microscopic model and it holds only in the limit of very weak inhomogeneities: higher order corrections, such as diffusion, will introduce a length scale, as we discuss in Section \ref{sec_force_second_order}.

Let us now revert to the more technical task of deriving the GHD equations \eqref{eq_ghd_1ord}. In the simpler case (\ref{Case1}) the equations can be proven in continuous integrable models where ultra-local conserved charges are a complete basis \cite{SciPostPhys.2.2.014}. The derivation needs the expectation value of certain generalized currents, which has now been microscopically derived \cite{PhysRevX.10.011054,10.21468/SciPostPhys.8.2.016,PhysRevLett.125.070602}. The hydrodynamics due to the second type of inhomogeneities (\ref{Case2}) can be proven only in certain instances, which we now discuss, and only conjectured out of these. Nevertheless, the validity of Eq. \eqref{eq_f2} has been benchmarked with iTEBD simulations \cite{PhysRevLett.123.130602}. More precisely, the case (\ref{Case2}) can be proven in spatially homogeneous, but time-dependent setups and in the inhomogeneous case in the presence of relativistic invariance. Apart from this case, the GHD equation \eqref{eq_ghd_1ord} is an educated guess based on a self-consistency requirement of being able to recast Eq. \eqref{eq_ghd_1ord} in the equivalent form for the filling \cite{PhysRevLett.123.130602}.

Albeit the case (\ref{Case1}) \cite{SciPostPhys.2.2.014} has been historically addressed before of (\ref{Case2}) \cite{PhysRevLett.123.130602}, for the sake of clarity we prefer to proceed in the opposite direction.
\ \\

\textbf{Derivation of the GHD equations for case (\ref{Case2})---}
Let us start considering a homogeneous system, but with a slowly time-dependent Hamiltonian $\alpha(t,x)\to \alpha(t)$: spatial inhomogeneities will be reintroduced afterwards.

We can approximate the slow change of the coupling $\alpha(t)$ with a staircase function, where at regular time intervals $\Delta t$ the coupling is changed of $\Delta \alpha$. Since we are assuming adiabatic changes, we first take the limit $\Delta t\to \infty$ and, only after, $\Delta \alpha\to 0$.
In this approximation, the time evolution is described by a sequence of infinitesimal quenches: \emph{i)} the system is initialized in a certain GGE build on the Hamiltonian $H(\alpha)$. \emph{ii)} Then one weakly excites the system changing $\alpha\to \alpha +\Delta \alpha=\alpha+\partial_t\alpha\Delta t$ and \emph{iii)} waits for a long time $\Delta t$ until relaxation to a new GGE takes place. Our task is now to connect the two GGEs before and after the infinitesimal quench and we can do it from the charges. Let $Q_i(\alpha+\Delta \alpha)$ be a conserved charge of the post quench Hamiltonian, then we have to require that its expectation value on the post quench GGE matches the initial one $\langle Q_i(\alpha+\Delta \alpha)\rangle_{t+\Delta t}=\langle Q_i(\alpha+\Delta\alpha)\rangle_t$. The left hand side is simply the expectation value of a charge on a GGE 
\be\label{eq_Qexp}
L^{-1}\langle Q_i(\alpha+\Delta \alpha)\rangle_{t+\Delta t}=\int \dd\lambda\, h_i^{(\alpha+\Delta \alpha)}(\lambda) \rho_{t+\Delta t}(\lambda)\, ,
\ee
where we made the $\alpha-$dependence of the charge eigenvalue explicit. Above, $L$ is the system's size, which appears for extensivity reasons. The second term is less trivial, since the state $\langle...\rangle_t$ is a GGE built on the Hamiltonian $H(\alpha)$ for which $Q_i(\alpha+\Delta\alpha)$ is \emph{not} a conserved quantity.
One can expand for small $\Delta \alpha$ as $\langle Q_i(\alpha+\Delta \alpha)\rangle_t\simeq\langle Q_i(\alpha)\rangle_t+\Delta\alpha\langle \partial_\alpha Q_i(\alpha)\rangle_t$: the expectation value $\langle Q_i(\alpha)\rangle_t$ is computed analouglsy to Eq. \eqref{eq_Qexp}, while the second term $\langle \partial_\alpha Q_i(\alpha)\rangle_t$ can be exactly obtained through a proper generalization of the Hellmann-Feynman (HF) theorem \cite{PhysRevLett.123.130602}.
In its standard formulation, the HF theorem considers an eigenvector of the Hamiltonian $H|\Psi\rangle=E|\Psi\rangle$ and relates energy variations with the expectation value of the derivative of the Hamiltonian $\partial_\alpha E=\langle \Psi|\partial_\alpha H|\Psi\rangle$. This identity only requires $|\Psi\rangle$ to be an eigenvector of $H$ and can be promptly generalized to other conserved quantities.
In our specific case, one obtains \cite{PhysRevLett.123.130602}
\be\label{eq_HF}
L^{-1}\langle \partial_\alpha Q_i(\alpha) \rangle_t =\int \dd\lambda\,\Big\{ \partial_\alpha h_i^{(\alpha)}(\lambda)\rho_t(\lambda)+\frac{f^\text{dr}(\lambda)}{(\partial_\lambda p)^\text{dr}}\partial_\lambda h_i^{(\alpha)}(\lambda) \rho_t(\lambda)\Big\}\, .
\ee
This result can now be combined with Eq. \eqref{eq_Qexp} (which we expand up to first order in $\Delta t$) and plug into the charge conservation, obtaining integral equations functions of the charge eigenvalue $h_i(\lambda)$
\be\label{eq_intchc}
\int \dd\lambda\, h_i^{(\alpha)}(\lambda) \partial_t\rho_{t}(\lambda)=
\partial_t\alpha\int \dd\lambda\, \frac{f^\text{dr}(\lambda)}{(\partial_\lambda p)^\text{dr}}\partial_\lambda h_i^{(\alpha)}(\lambda) \rho_t(\lambda)\, .
\ee
Notice that the $\propto \partial_\alpha h_i^{(\alpha)}(\lambda)$ term in Eq. \eqref{eq_HF} exactly balances the similar term coming from the expansion of Eq. \eqref{eq_Qexp}. Therefore, when charge conservation is imposed, $\partial_\alpha h_i^{(\alpha)}$ does not appear in the expression. We can now integrate by parts the r.h.s. of Eq. \eqref{eq_intchc} and, assuming boundary terms do not contribute \footnote{This assumption is not always correct and can lead to unexpected consequences. For example, in the XXZ spin chain in the planar regime boundary terms cannot be neglected, as we discuss in Section \ref{sec_hidd_nonadiab}.}, write the above integral equation as $\int \dd\lambda\, h_i^{(\alpha)}(\lambda)\{ \partial_t\rho_{t}(\lambda)+
\partial_t\alpha \partial_\lambda [\frac{f^\text{dr}(\lambda) }{(\partial_\lambda p)^\text{dr}}\rho_t(\lambda)]\}=0$.
Then, in the last step one invokes the completeness of the charges and imposes the integral equations hold for arbitrary functions $h^{(\alpha)}_i(\lambda)$ and the desired equation is reached $\partial_t\rho_{t}(\lambda)+
\partial_t\alpha \partial_\lambda [\frac{f^\text{dr}(\lambda)}{(\partial_\lambda p)^\text{dr}}\rho_t(\lambda)]=0$, albeit in the simpler homogeneous case.

We now show how, in the case of relativistic invariance, from the homogeneous GHD equation we can recover the more general form. Let us start considering the inhomogeneous GHD equations for the filling \eqref{eq_ghd_1ord_filling} and in the absence of the force term, which we write in the following form $(\partial_\lambda p)^\text{dr}\partial_t n+(\partial_\lambda \epsilon)^\text{dr}\partial_x n=0$. In the relativistic case, this equation can be brought into an explicit covariant continuity equation. Indeed, if one assumes a relativistic dispersion law $\epsilon(\lambda)=m\cosh\lambda$ and $p(\lambda)=m\sinh\lambda$, with $m$ the mass of the excitation, it holds $\partial_\lambda \epsilon=p$ and $\partial_\lambda p=\epsilon$. One then collects energy and momentum in a quadrivector $P^\mu$ s.t. $P^0=\epsilon$ and $P^1=p$ and rewrites the GHD equations as $(P^\mu)^\text{dr}\partial_\mu n=0$ (sum over repeated indexes), with $\partial_\mu=(\partial_t,\partial_x)$. The continuity equation is explicitly Lorentz invariant: the dressing operation does not change the transformation properties of the quadrimomentum. This can be easily checked using that the filling is Lorentz invariant. Now, let us go back to the force term: the homogeneous equations we previously derived can be written as $\epsilon^\text{dr}\partial_t n+\partial_t\alpha f^\text{dr}\partial_\lambda n=0$. This equation must be the restriction to the homogeneous case of a Lorentz-invariant one
\be
(P^\mu)^\text{dr}\partial_\mu n+\partial_\mu \alpha (\mathcal{F}^\mu)^\text{dr} \partial_\lambda n=0\, ,
\ee
with $\mathcal{F}^0(\lambda)=f(\lambda)$ and $\mathcal{F}^1(\lambda)=\Lambda(\lambda)$. Imposing that $\mathcal{F}^\mu$ behaves as a quadrivector under Lorentz boosts, the form of $\Lambda(\lambda)$ is completely determined by the knowledge of $f(\lambda)$, resulting in Eq. \eqref{eq_f2} (restricted to the Lorentz-invariant case).
\ \\

\textbf{Derivation of the GHD equations for case (\ref{Case1})---} The GHD equations can be derived for the case (\ref{Case1}) without the need of invoking Lorentz invariance. More precisely, the argument provided in Ref. \cite{SciPostPhys.2.2.014} also assumes a continuum theory and the completeness of the local charges (i.e. the charge density ${\rm q}_j(x)$ is expressed in terms of the fields and a finite number of their derivatives computed in $x$).
Let us consider an inhomogeneity in the form
\be
H=\int \dd x\, {\rm h}(x)+\alpha(t,x){\rm q_j}(x),
\ee
where ${\rm h}(x)$ and ${\rm q}_j(x)$ are the Hamiltonian and an ultralocal charge density of the model in the absence of inhomogeneity.
We now consider the time evolution of a charge density ${\rm q}_k(x)$. In the Heisenberg picture one has
\be\label{eq_chev}
\partial_t {\rm q}_k(x)=i\Big[\int \dd y\, {\rm h}(y)+\alpha(t,y){\rm q_j}(y),{\rm q}_k(x)\Big]=-\partial_x {\rm j}_k+i\int \dd y\, \alpha(t,y)[{\rm q_j}(y),{\rm q}_k(x)]\, .
\ee
Above, we used that in the absence of the inhomogeneity, the time derivative of the charge is the gradient of the associated current ${\rm j}_k(x)$.
Following Ref. \cite{SciPostPhys.2.2.014}, one can expand the commutator of the two charge densities in a local basis of operators
\be\label{eq_2chdc}
i[{\rm q_j}(y),{\rm q}_k(x)]=\sum_\ell \delta^{(\ell)}(y-x) O_{j,k}^{\ell}(x)\, .
\ee
Above, $\delta^{(\ell)}$ is the $\ell^\text{th}-$derivative of the Dirac delta function. Using this representation in Eq. \eqref{eq_chev}, we express the evolution of the charge as a derivative expansion of the inhomogeneity

\be\label{eq_chev1}
\partial_t {\rm q}_k(x)=-\partial_x {\rm j}_k+\sum_\ell (-\partial_x)^\ell[\alpha(t,x)O^{\ell}_{j,k}(x)]\,.
\ee
Since one is interested only in the first corrections due to the inhomogeneity, only the $\ell=0$ and $\ell=1$ terms in the sum are retained. Now, we determine the $O_{j,k}^\ell$ operators of interest. Integrating Eq. \eqref{eq_2chdc} over the $y$ coordinate, one gets the identity $O^0_{j,k}(x)=i[Q_j,{\rm q}_k(x)]$: the commutator can be interpreted as the time derivative of the Heisenberg equation of motion for the charge density ${\rm q_k}(x)$ which evolves with the Hamiltonian $Q_j$. As a consequence, the commutator is nothing but the derivative of a proper current: let us call ${\rm j}_k^j$ the generalized current associated to the charge density ${\rm q}_k$ under the evolution generated by $Q_j$.
Hence, $O^0_{j,k}(x)=i[Q_j,{\rm q}_k(x)]=-\partial_x {\rm j}_k^j(x)$.
Repeating the same reasoning in Eq. \eqref{eq_chev}, but integrating over $x$ rather than $y$, one reaches
\be
\partial_y {\rm j_i^k}(y)=\sum_\ell (-\partial_y)^\ell O_{j,k}^\ell(y)\, .
\ee
Using $O^0_{j,k}(y)=-\partial_y {\rm j}_k^j(y)$, one can arrange the equation above as an identity for $\partial_y O^1_{j,k}(y)$ which, after integrating over $y$, becomes
\be
 O^1_{j,k}(y)=- {\rm j}_j^k(y)- {\rm j}_k^j(y)+A\mathbbm{1}+\sum_{\ell\ge 2}(-\partial_y)^{\ell-1} O^{\ell}_{j,k}
(y)\, .
\ee
Above, $A$ is an integration constant that cannot be fixed with these considerations. Based on transformations under parity and UV arguments, one can argue $A=0$: we leave these considerations to the original reference \cite{SciPostPhys.2.2.014}.
We can now go back to Eq. \eqref{eq_chev}, use the expressions for $O_{j,k}^0$ and $O_{j,k}^1$ and retain up to the first order derivatives
\be
\label{eq:qjjj}
\partial_t {\rm q}_k(x)=-\partial_x \big({\rm j}_k+\alpha(t,x){\rm j}_k^j\big) -\partial_x \alpha(t,x){\rm j}_j^k(x)+...
\ee
The last step consists in taking the expectation values of the above identity under the assumption that the system is locally described by a GGE, then invoke the completeness of the set of charges to get the final hydrodynamic equation. These operations are analogue to what we did for the derivation of Case (\ref{Case2}) and therefore will not be reported. The expectation value of ${\rm j}_k^j$ is a simple generalization of the usual current
\be
\label{eq:currgge}
\langle {\rm j}_k^j\rangle= \int \dd\lambda\, h_k(\lambda)\frac{(\partial_\lambda h_j(\lambda))^\text{dr}}{(\partial_\lambda p(\lambda))^\text{dr}}\rho(\lambda)\, .
\ee

The expression for the expectation value of the currents has been proposed as an educated guess in the first GHD references \cite{GHD1,GHD2}, lately generalized to the flow of arbitrary charges in Ref. \cite{SciPostPhys.2.2.014}. More recently, a rigorous proof have been presented
\cite{PhysRevX.10.011054,10.21468/SciPostPhys.8.2.016,PhysRevLett.125.070602}.
This concludes the derivation of the first-order hydrodynamic equations. We now move to discuss their application to inhomogeneous setups, in particular trapped systems.

\subsection{Relaxation in trapping potentials and lack of thermalization}
\label{sec_firstorder_lackth}

A relevant application of the GHD equation with force fields of type (\ref{Case1}) is related to the presence of confining potentials or \textit{traps}. This is particularly important because of its connection with experimental protocols, as the pioneering experiment dubbed ``Quantum Newton cradle"~\cite{Kinoshita2006}. In several cases, the interaction between bosonic atoms takes the form of a contact one, which leads to the Hamiltonian
\begin{align}\label{eq_LL_H}
&H = H_{\rm LL} + \int dx \, V(x) \psi^\dag(x) \psi(x) \;,\\ &H_{\rm LL} =  \int dx\, \left[\frac{\hbar^2}{2m}\partial_x \psi^\dag(x) \partial_x \psi(x)  + \intc \, \psi^\dag(x) \psi^\dag(x) \psi(x) \psi(x)\right],
\end{align}
where $H_{\rm LL}$ is the Lieb-Liniger (LL) Hamiltonian \cite{PhysRev.130.1605,PhysRev.130.1616} and the potential term $V(x)$ accounts for the confining effect of the trap, which can be expanded as a harmonic potential plus possibly anharmonic corrections, e.g. $V(x) = \frac 1 2 m \omega^2 x^2  + o(x^2)$.
As it is standard, it is convenient to work in units such that $\hbar^2/2m=1$. The LL model within the repulsive phase $c>0$ features a single species of excitation, which experience the following scattering phase
\be\label{eq_ScM_LL}
\Theta(\lambda)=-2\arctan(\lambda/c)\, .
\ee
Within the attractive phase $c<0$, the model sustain bound states, as we extensively discuss in Section \ref{eq_sec_attractive_bose_gas}, but within this section we focus on the repulsive regime.
As in the standard quantum quench protocol, the system is prepared in an out-of-equilibrium state and let evolve with the unitary dynamics generated by $H$. For instance in \cite{Kinoshita2006}, an initially thermal state was put out-of-equilibrium by the action of a Bragg-pulse sequence which splits the atoms into two clouds in momentum space.

\begin{figure}
    \centering
    \includegraphics[width=0.9\textwidth]{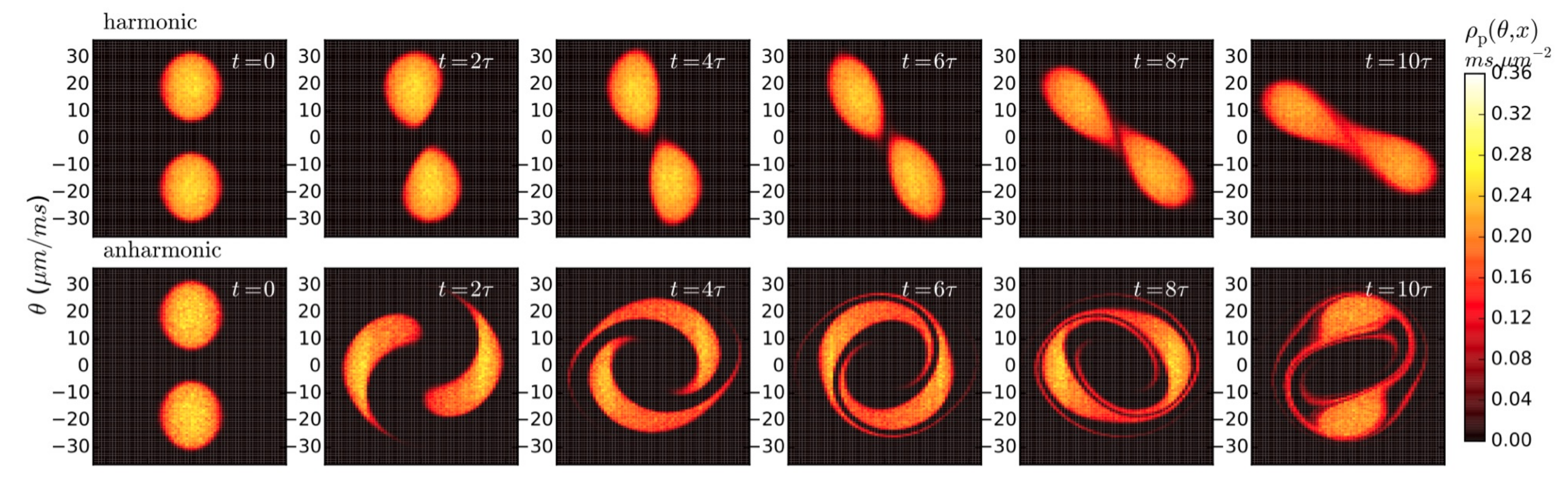}
    \caption{From \cite{10.21468/SciPostPhys.6.6.070}: Euler Evolution~\eqref{eq_ghd_1ord} in the phase space $(x, \theta)$ of the root density $\rho_{t,x}(\theta)$ of the Lieb-Liniger model for a harmonic potential (top) and anharmonic one (bottom);  $\tau$ is the trap period (see~\cite{10.21468/SciPostPhys.6.6.070} for details). The initial state is obtained by joining identical thermal distributions boosted in opposite directions. }
    \label{fig:llharmoanharmo}
\end{figure}

Beyond the non-interacting limits ($c=0$ -- free bosons, and $c \to\infty$ -- Tonks-Girardau), the exact theoretical description of this kind of procedures is a formidable task because of the presence of strong interactions and explicit integrability-breaking terms due to the trap. However, for initial configurations and potentials 
sufficiently smooth in space, one can apply the technology of GHD described in the previous section. Once an approximate description of the initial state in terms of a (quasi)-stationary local GGE state has been obtained~(see for instance the discussion in \cite{PhysRevLett.122.090601, 10.21468/SciPostPhys.6.6.070}), one can follow the evolution in space and time by numerically solving the first order equation, Eq.~\eqref{eq_ghd_1ord}. This approach has provided accurate theoretical predictions in very good agreement with experimental data~\cite{PhysRevLett.122.090601, malvania2020generalized} and microscopic simulations~\cite{PhysRevLett.120.164101, PhysRevLett.125.240604}. A characteristic example from \cite{10.21468/SciPostPhys.6.6.070} is shown in Fig.~\ref{fig:llharmoanharmo}.

A basic question regards the large-time behavior of this dynamical procedure. Does a stationary state emerge? If so, can this state be described by a thermal ensemble? 
A first observation is related to the role of interactions. Consider as a simple example the non-interacting limit $c=0$ and a harmonic potential $V(x) = \frac{1}{2} m \omega^2 x^2$. In this case, all particles undergo independent oscillations with the same frequency $\omega$: as a consequence, the cloud as a whole rigidly rotates in the phase space with perfect periodicity and no stationary state can emerge. In contrast, when interactions are turned on (see Fig.~\ref{fig:llharmoanharmo} -- top), different portions of the phase space rotate at a different frequency because of the effect of dressing in \eqref{eq_ghd_1ord}. In the presence of anharmonicity, the distribution is stirred up by time-evolution even in the absence of interactions \cite{PhysRevB.95.174303}: in this case, interactions and dephasing can merge different filaments~(see Fig.~\ref{fig:llharmoanharmo} -- bottom).
This kind of deformation in phase space becomes more and more evident at larger times: while the root density remains smooth under Eq.~\eqref{eq_ghd_1ord}, it becomes rugged at a finer and finer scale. 

One can see that this evolution does not have a \textit{point-wise} limit for the root density $\rho_{t,x}(\lambda)$ at large times. In fact, even in the presence of an integrability-breaking potential, the first order equation \eqref{eq_ghd_1ord} still has a large number of conserved quantities. Indeed, for any weakly differentiable function $f$, we can define
\begin{equation}
\label{eq:fillingcharge}
    \mathcal{Q}_f = \int d\lambda dx \; f\bigl(n(\theta)\bigr) \rho_{t,x}(\theta),
\end{equation}
and deduce from \eqref{eq_ghd_1ord} and \eqref{eq_ghd_1ord_filling}
\begin{multline}\label{eq_dQdt}
\partial_t \mathcal{Q}_f  = \int d\lambda \, dx \; [ \rho f'(n) \partial_t n + f(n) \partial_t \rho] = \\ = 
- \int d\lambda dx \; [ f'(n) (v^\text{eff} \partial_x n + F^\text{eff} \partial_\lambda n) + f(n) (\partial_x(v^\text{eff} \rho) + \partial_\lambda(F^\text{eff} \rho)] =\\ = -\int d\lambda dx \; \left[\partial_x (v^\text{eff} f(n) \rho) + \partial_\lambda (F^{\rm eff} f(n) \rho) \right]=0
\end{multline}
The last integrand is manifestly a pure divergence and thus the integral vanishes via Stokes' theorem whenever boundary terms can be neglected~(see Sec.~\ref{sec_hidd_nonadiab} about this). Although the integrability of the GHD equation has been recently analysed~\cite{DOYON2018570, 2017JPhA...50Q5203B}, its validity in the presence of a force field and the nature of the large set of conserved quantities in \eqref{eq:fillingcharge} is not yet fully understood. However, choosing for any $n_0 \in [0,1]$ the function $f(n) = \delta(n-n_0)$, one can see that $\mathcal{Q}_f$ measures the volume of the phase space corresponding to a given value $n_0$ of the filling fraction. Additionally, for the choice
\begin{equation}\label{eq_entropy}
f(n) = \frac{1}{n} ( - n \ln n - (1-n) \ln (1-n)),
\end{equation}
$\mathcal{Q}_f$ coincides with the Yang-Yang entropy and its conservation implies the absence of entropy production under the GHD first order dynamics~\cite{10.21468/SciPostPhys.6.6.070}. 

Nevertheless, while point-wise convergence of $\rho_{t,x}(\lambda)$ at large times is not to be expected, the functional dependence of a local observable $O[\rho]$ in only sensitive to a coarse-grained version $\bar{\rho}_{t,x}$, corresponding to a local average in a small phase-space cell of the original root  density, i.e.
\begin{equation}
\label{eq:coarsegrained}
    O[\rho] \sim O[\bar \rho] \;, \qquad \bar \rho_{t,x}(\lambda) = \frac{1}{4\delta x \delta \lambda}\int_{\lambda - \delta \lambda}^{\lambda + \delta \lambda} d\lambda' 
    \int_{x - \delta x}^{x + \delta x} dx' \rho_{t,x'}(\lambda')
\end{equation}
As argued in \cite{10.21468/SciPostPhys.6.6.070}, the first order GHD flow is invariant under the replacement $\rho \to \bar \rho$: the evolution of a coarse grained initial condition coincides with the coarse-graining of the evolved root density. This is particularly important because any numerical procedure
will necessarily involve some form of coarse-graining: an ultraviolet cutoff naturally emerges from the lattice discretization of \eqref{eq_ghd_1ord} or from microscopic interparticle separation in a molecular dynamics simulation~\cite{PhysRevLett.120.045301}.

Note that the smoothening procedure in \eqref{eq:coarsegrained} generally breaks the conservation laws in Eq.\eqref{eq:fillingcharge} and leads in particular to the production of Yang-Yang entropy. 
One can thus wonder whether there exists a large-time stationary limit to the coarse grained root density $\bar{\rho}_{t,x}(\lambda)$ and what is its nature. Extensive numerical analysis have shown that a stationary limit does indeed emerge though this resulting state is not thermal~\cite{10.21468/SciPostPhys.6.6.070, PhysRevLett.120.164101}.  This suggests a residual integrability for the coarse-grained GHD first order evolution also in the presence of inhomogeneous force terms. A clear intuition can be gained by turning at the non-interacting case ($\intc = 0$) in the presence of an anharmonic potential~\cite{PhysRevB.95.174303}. In this case, every point in the phase space belongs to a closed orbit with a different period and a stationary state emerges because of dephasing. The integral of the root density along each orbit provides a set of conserved quantities which fully characterizes the stationary state. It remains an open problem to extend this picture in the presence of interactions.

\subsection{Diffusive corrections and inhomogeneity-induced thermalization }
\label{sec_force_second_order}
After having discussed the existence and the characterization of the stationary state for the first order evolution \eqref{eq_ghd_1ord}, it is natural to investigate how the higher order corrections to the hydrodynamic description affect the large time behavior. For purely integrable Hamiltonians (i.e. no space-time dependent parameters as in \eqref{eq:halpha}), the higher order corrections to GHD can be obtained via the \textit{constitutive relations}, which relates the expectation value of the generalized currents at the point $x$ to the derivatives of all charges around $x$ itself
\begin{equation}
\label{eq:currdiff}
    j_i^k(x) = \mathcal{F}_i^k(\{q(x)\})  - \frac 1 2 \sum_{j} \mathcal{D}_i^{jk} (\{q(x)\}) \partial_x q_j(x) + O(\partial_x^2 q_j(x)) \equiv  [j_i^k(x) ]_{\rm homo}\; .
\end{equation}
Conventionally, we will simply omit the index $k$ when referring to the standard Hamiltonian evolution. In \eqref{eq:currdiff}, the first term only involves expectation values \eqref{eq:currgge} on top of stationary GGE states  and leads to the standard first order Eulerian GHD. The second one is responsible for the diffusive corrections as can be seen plugging \eqref{eq:currdiff} in the continuity equation \eqref{eq_chj_continuity}
\begin{equation}
\label{eq:chargediff}
\partial_t q_i(x,t) + \partial_x \mathcal{F}_i (q (x,t)) = \frac12 \sum_j \partial_x [\mathcal{D}_i^j (\{q( x)\}) \partial_x q_j( x)]\, .
\end{equation}
The explicit form of the diffusion matrix $\mathcal{D}_i^j$ was computed in \cite{PhysRevLett.121.160603,10.21468/SciPostPhys.6.4.049}, by relating it to the current-current correlator; in particular, assuming that all conserved quantities can be chosen as PT-symmetric, one has the Einstein relation (see e.g.~\cite{10.21468/SciPostPhys.6.4.049}) 
\begin{equation}
\label{eq:LDrel}
    \mathcal{L}_{ij} = \sum_k \mathcal{D}^k_i C_{kj},
\end{equation}
where we introduced the Onsager matrix $\mathcal{L}_{ij}$, which is expressed in terms of the susceptibilities $C_{ij}$ and the Drude weight $D_{ij}$~\cite{PhysRevB.97.081111, PhysRevLett.119.020602, GHDDrudeBoseGas, PhysRevLett.82.1764, PhysRevLett.119.080602} defined as follows
\begin{subequations}
\label{eq:transpdef}
\begin{align}
&    C_{ij} \equiv  \int dx \, \langle \qq_i(x,0) \qq_j(0,0) \rangle^c  
\label{eq:suscC}
\\
&    D_{ij} \equiv \lim_{t \to \infty} \frac 1 {2t} \int_{-t}^t ds  \int dx \langle \jj_i(x,s) \jj_j(0,0) \rangle^c \label{eq:Drude} \\
&    \mathcal{L}_{ij} \equiv \lim_{t\to \infty} \left(\int_{-t}^t ds\int dx \, \langle \jj_i(x,s) \jj_j(0,0) \rangle^c - D_{ij}\right) \label{eq:Ldef}
\end{align}
\end{subequations}
and $\langle \ldots \rangle^c$ indicates the connected correlator. The definitions \eqref{eq:transpdef} are generally applicable in the presence of a set (finite or infinite) of conserved quantities. A few comments are in order here: i) The susceptibility matrix \eqref{eq:suscC} on a GGE $e^{-\sum_j\beta_j Q_j}$ can also be expressed as $C = \partial q_i / \partial \beta_j $, and thus appears naturally when passing from the generalised inverse temperatures $\beta_j$ to the expectation value of conserved densities (canonical to microcanonical); ii) A non-zero Drude weight $D$ indicates the presence of ballistic transport in the model; it is thus an important transport quantity in integrable systems, which the Kubo formula \eqref{eq:Drude} expresses as a property of a thermodynamic state\footnote{We stress that although the same letter $D$ and $\mathcal{D}$ are generally used respectively for the Drude weight and the diffusion constant, they are very different quantities which quantify the ballistic and diffusive components.}; iii) 
the Onsager matrix $\mathcal{L}$ defined by \eqref{eq:Ldef} can be related to the diffusion matrix \eqref{eq:LDrel} using the constitutive relation \eqref{eq:currdiff} and its consequences, via linear response, on two-point functions~\cite{Doyon_2017, 10.21468/SciPostPhys.6.4.049}. Physically it corresponds (up to trivial temperature-dependence factors) to the conductivity of the system.

In integrable systems, using the quasiparticle representation, the current-current correlator involved in \eqref{eq:Ldef} has been obtained generalizing the exact thermodynamic form factors of the density operator in the Lieb-Liniger model for two particle-hole excitations~\cite{De_Nardis_2018}. 
We refer the reader to the chapter about correlation functions and transport coefficients in this volume. 
In the general case of inhomogeneous couplings in the Hamiltonian, the derivation of a full GHD equation is technically challenging and only recently a hydrodynamic description has been obtained~\cite{2021arXiv210513068D}.
The motivation behind this is that inhomogeneous couplings lead to an explicit breaking of integrability (see Sec.~\ref{sec:diffhydr}) which must be combined with the already known GHD diffusion induced by the state inhomogeneity~\eqref{eq:chargediff}. More concretely, consider the Hamiltonian 
\begin{equation}
\label{eq:hpot}
    H = H_0 + \int dx \, \pot(x) \qq_{\barj}(x)
\end{equation}
coupled to a specific conserved density of index $\barj$. We
expand the inhomogeneity around the center $\bar x$ of a hydrodynamic cell
\begin{equation}
\label{eq:pottay}
        \pot(x) = \pot(\bar x) + (x-\bar x) \partial_x \pot(\bar x) + \frac{1}{2}(x-\bar x)^2 \partial_x^2 \pot(\bar x) + O(\partial_x^3 \pot) 
\end{equation}
and to simplify the notation, we set $\bar x= 0$ in the following.
Then, up to second derivatives, a first correction appears as the $\ell = 2$ term in \eqref{eq_chev}. However, second derivatives $\partial_{x}^2 \alpha(x,t)$ could enter the evolution of a charge density $\qq_k(\bar x)$ only coupled with 
    \begin{equation}
        \int dx x^2  \langle [\qq_{\barj}(x), \qq_k(0)]\rangle_{\rm GGE} = 0\, ,
    \end{equation}
which vanishes whenever PT invariance holds. 
More importantly, the constitutive relation \eqref{eq:currdiff} can be modified as
\begin{equation}
    \label{eq:currdiffinh}
    j_i^k(x) = 
     [j_i^k(x)]_{\rm homo} +  
    \Omega_{\barj; i}^{k} (\{q(x)\}) \partial_x \alpha(x) + \ldots\;,
\end{equation}
where $[j_i^k(\bar x)]_{\rm homo}$ contains the homogeneous contributions as in \eqref{eq:currdiff} and $\Omega_{\barj; i}^{k}(\{q(\bar x)\})$ encodes the coupling to the potential. In principle, one could plug Eq.~\eqref{eq:currdiffinh} in \eqref{eq:qjjj}, but the computation of the coefficients $\Omega_{\barj; i}^{k}$ is difficult in general as they cannot be fully expressed in terms of two particle-hole form factors~\cite{2021arXiv210513068D}. We refer the reader to the original reference. In this review, we will focus on the special case for which $\Omega_{\barj; i}^{k}$ can be shown to vanish identically for all values of $i, k$. This happens when the conserved density $\qq_{\barj}(x)$ in Eq.~\eqref{eq:hpot} has a current which is itself a conserved density
\begin{equation}
\label{eq:conscurr}{}
    \partial_t \qq_{\barj} + \partial_x \jj_{\barj}= 0 \;, \qquad \jj_{\barj} \equiv \qq_{\barj + 1}\, .
\end{equation}
This can be the consequence of Galileian  (e.g. the current associated to particle number is the momentum density) or Lorentz invariance (e.g. the energy current is the momentum density). More generally, it is implied by the existence of the boost operator~\cite{GRABOWSKI1995299, 10.21468/SciPostPhys.9.3.040, PhysRevB.90.161101}. Such a conserved density has \textit{vanishing diffusion}: using that the integrated current $\int dx \; \jj_\barj(x)$ is time independent in Eqs.~(\ref{eq:Drude}, \ref{eq:Ldef}), it follows from \eqref{eq:LDrel} that 
\begin{equation}
\label{eq:nodiff}
\mathcal{L}_{\barj i}^k  =    \mathcal{D}_{\barj}^{ik} = 0 \;, \qquad \forall i,k
\end{equation} 
In this case, starting from an equilibrium state at a reference time $t = 0$, we see that $\Omega_{\barj; i}^k$ will be generated time-evolving with \eqref{eq:hpot} the current operator $\jj_i^k$. Thus, using \eqref{eq:pottay}
\begin{align}
    \Omega_{\barj; m}^{k} &\sim  \imath \int_0^t ds \int dx' \; x' \langle[\qq_\barj(x', s), \jj_m^k(x, t)]\rangle_{\rm GGE} =\\ 
    & - \imath \int_0^t ds \; s \int dx' \; x' \langle[\partial_s \qq_\barj(x',s), \jj_m^k(x, t)]\rangle_{\rm GGE} =\\
    & \imath \int_0^t ds \; s \int dx' \; x' \langle[\partial_x \jj_\barj(x',s), \jj_m^k(x, t)]\rangle_{\rm GGE} \, ,
\end{align}
where in the second line we used integration by parts and in the third line 
the continuity equation \eqref{eq_chev} up to $\ell = 0$. Using again integration by parts with respect to $x'$ and that the integrated current is a conserved quantity, we arrive at $\Omega_{\barj; m}^{k} = 0$.

Paradigmatic examples where \eqref{eq:conscurr} holds are the density operator $\qq_\barj(x) \to \psi^\dag(x) \psi(x)$ in the Lieb-Liniger model, or the energy density $\qq_\barj(x\sim n) \to s^x_n s^x_{n+1} + s^y_n s^y_{n+1} + \Delta s^z_n s^z_{n+1}$ for the Heisenberg XXZ chain. 
We focus on the former which is the most relevant for experimental applications and satisfies an additional technical simplification: the density $\qq_{\barj}(x)$ \textit{does not generate any evolution flow}, i.e.
\begin{equation}
\label{eq:noflow}
    \int dx [\qq_{\barj}(x), \qq_k (y)] = 0 \;, \qquad \forall k
\end{equation}
and this implies that $F_k^{\barj} = \mathcal{D}_{k}^{i\barj} = 0$ for all $k$ and $i$, and in the following we will make this assumption.
Then, the diffusive GHD equation in the presence of an inhomogeneous potential \eqref{eq:hpot} satisfying (\ref{eq:conscurr}, \ref{eq:noflow}) can be obtained by adding the first order force term $O_{j,k}^{\ell = 1} \partial_x \alpha$ to the already known diffusive GHD equation \eqref{eq:chargediff}, i.e.
    \begin{equation}
\label{eq:chargediffinho}
\partial_t q_i + \partial_x \mathcal{F}_i (q ) = \frac12 \sum_j \partial_x [\mathcal{D}_i^j (\{q\}) \partial_x q_j] -  
\mathcal{F}_{\barj}^i (q )\, \partial_x \alpha.
\end{equation}
This equation can be efficiently solved numerically (see \ref{sec_num}) and gives access to an accurate description of the out-of-equilibrium dynamics up to the diffusive order~\cite{PhysRevLett.125.240604}. 

Also, important consequences can be deduced analysing the stationary state. Contrarily to the first order dynamics, the diffusive term in Eq.~\eqref{eq:chargediffinho} leads to the production of entropy, which can be compactly written in terms of the spatial derivatives of the GGE Lagrange multipliers $\beta_j(x)$ conjugated to the charge densities $\qq_j(x)$ as ~\cite{10.21468/SciPostPhys.6.4.049, 2019arXiv191201551D, 10.21468/SciPostPhysLectNotes.18}
\begin{equation}\label{eq_dtS}
\frac{dS}{dt} = \frac 12 \sum_{jk} \int dx \; [\partial_x \beta_j]\, \mathcal{L}_{jk}\, [\partial_x \beta_k ] \;.
\end{equation}
Positivity of the Onsager matrix can be proven~\cite{10.21468/SciPostPhys.6.4.049, 10.21468/SciPostPhysLectNotes.18} and it implies an increase of entropy. At stationarity, the entropy must saturate and combining this with \eqref{eq:nodiff}, we arrive at the condition
\begin{equation}
    \label{eq:noentro}
    \partial_x \beta_k = 0  \;, \quad \forall k\neq \barj\, .
\end{equation}
Within this manifold of states, the diffusion constant $\mathcal{D}_i^j$ in \eqref{eq:chargediffinho} vanishes and the dynamics is controlled by the Eulerian part. Imposing stationarity of the Eulerian evolution within the manifold \eqref{eq:noentro}, one deduces that, in the local density approximation up to the second derivative, the system relaxes to the standard Gibbs ensemble
density matrix of the form~\cite{PhysRevLett.125.240604}
\begin{equation}
\label{eq:thermalrho}
\varrho_{\rm Thermal} = \frac{1}{\mathcal{Z}} \exp[ -\beta (H - \mu  Q_\barj)]
\end{equation}
where $H$ is given in \eqref{eq:hpot}, while $\beta, \mu$ play the role of an inverse temperature and chemical potential, fixed respectively by the initial expectation values of $H$ and $Q_\barj$. Given that the thermal density matrix in \eqref{eq:thermalrho} is expected to be the true stationary state, it is believed that higher order corrections to GHD will not affect the long-time behavior.

Summarizing, we thus see that, starting from any smooth initial state, Eq.~\eqref{eq:chargediffinho} implies the existence of three time regimes, associated to the length scale $\ell$ of variation of the potential $\pot(x)$: 
\begin{enumerate}
\item at short time $t \ll  \ell / v^{\text{eff}} $, the system equilibrates to a local GGE state, characterized by the initial value of the conserved densities;
\item at times $\ell / v^{\text{eff}}  \lesssim t \ll \ell^2 / ||\mathcal{D}||_2$ (the norm $||\mathcal{D}||_2$ gives the typical scale of the diffusion) ballistic transport kicks in bringing to a pre-thermalisation plateau at the Eulerian stationary state described in Sec.~\ref{sec_firstorder_lackth};
\item finally at $t \gtrsim \ell^2 / ||\mathcal{D}||_2$, diffusive effects become important leading to full relaxation to a thermal state.
\end{enumerate}

For times $t\gtrsim \ell^2/ ||\mathcal{D}||_2$, corrections beyond the diffusive scale start to be non-negligible, however the thermal state has already been reached and is believed to be stationary for the exact dynamics. Thus, the GHD with diffusion is expected to describe the whole dynamics up to infinite times.

A word of caution should be spent about the relaxation to the thermal steady state.
The entropic argument Eq. \eqref{eq_dtS} shows that, if a steady state is reached, this must be thermal, but does not necessarily imply relaxation. This is the case, for example, of sloshing modes in perfectly harmonic traps in the 1d Bose gas. Indeed, displacing the center of the cloud with respect to the bottom of the parabola and then letting the system evolve, the center of mass undergoes persistent undamped oscillations at any time scale \cite{PhysRevA.72.063620}. However, this is a fluke of the harmonic potential for which the center of mass decouples from all the other degrees of freedom and is thus broken by any weak anharmonicity.

Lastly, we would like to comment on the expected thermalization time scales. Assuming the system is relaxing and thus thermalization is attained, the spectrum of the Onsager matrix determines the thermalization time scales through Eq. \eqref{eq_dtS}. Even though the Onsager matrix has only one zero eigenvalue associated with the $Q_\barj$ conservation, it still has arbitrarily small eigenvalues.
For this reason, the approach to a thermal state is typically controlled by a power law without a properly defined thermalisation time scale. So, in practice, the emergence of a thermal state is often hard to observe both numerically and experimentally~\cite{PhysRevLett.125.240604}.

\begin{figure}[ht]
         \centering
           \includegraphics[width=0.9\textwidth]{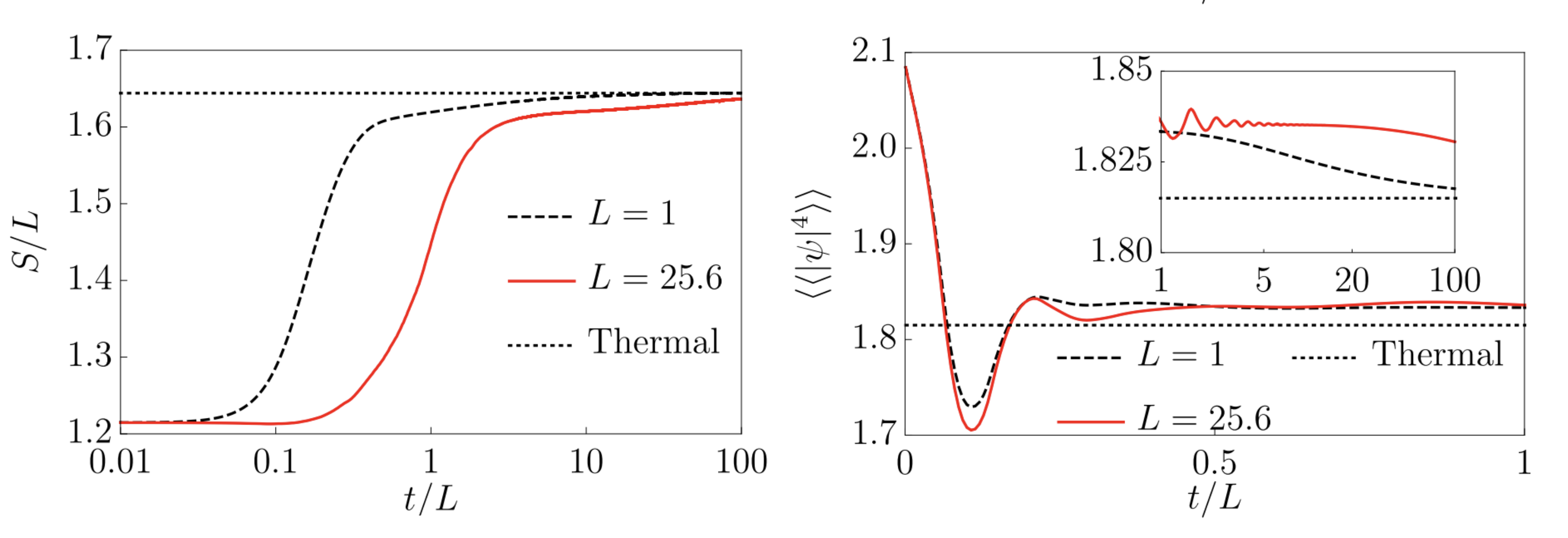}                            \caption{From \cite{PhysRevLett.125.240604}: The Lieb-Liniger model \eqref{eq_LL_H} initially prepared in a thermal state with $\beta = c = 0.3$ and chemical potential $\mu=2.5$ inside the potential $V_0(x) = 2 \sin(2 \pi x /L)$ is quenched by a sudden change of the potential to $V_0(x) \to V(x) = 1/2 ( 1 - \cos(2\pi x /L))$. Periodic boundary conditions are assumed. Larger values of $L$ correspond to a weaker diffusion and a longer plateau controlled by the Eulerian dynamics. Left: Time evolution  of the Yang-Yang entropy. Saturation to the thermal value is shown, whose timescale is controlled by the diffusion strength $L^{-1}$. Right: Four-point correlator $\langle \langle |\psi|^4 \rangle\rangle\equiv L^{-1}\int_0^L d x \langle \psi^\dagger(x)\psi^\dagger(x)\psi(x)\psi(x)\rangle$ as a function of time. At short time the Eulerian dynamics dominates and the evolution with scaled time $t/L$ becomes independent of $L$, relaxing to a  pre-thermal plateau. In the inset much larger times are shown and the drift towards the thermal value is clearly visible.  }
    \label{fig:thermdiff}
\end{figure}

\section{Generalized Boltzmann equation and collision integrals}
\label{sectionBoltzmann}

In this section, we discuss how integrability breaking can be implemented explicitly in the GHD framework (beyond the ``force'' terms discussed in the previous section) by introducing a {\em collision integral} term in the GHD equation~\cite{10.21468/SciPostPhys.6.6.070,PhysRevB.101.180302,2020arXiv200411030D}. The phrase {\em collision integral}  is borrowed from the Boltzmann equation terminology of weakly interacting particles, and will encode in a generic way diffractive scattering processes, quasiparticle decay, and other processes breaking integrability~\cite{PhysRevA.87.012707}. In what follows, we will first discuss the generic consequences of integrability breaking without specifying the form of the collision integral. In general, integrability breaking leads to diffusive hydrodynamics for the residual conserved charges~\cite{PhysRevB.101.180302}, in a way reminiscent of the Drude model of solid state physics. Quasiparticles move ballistically over a mean free path, but end up scattering due to integrability breaking, leading to a finite conductivity given by the Drude weight times the mean free time. The GHD formalism allows one to make this intuition precise, and to extract an exact expression for the way Drude weights are broadened into (generalized) Lorentzian upon breaking integrability~\cite{PhysRevB.101.180302,2020arXiv200411030D}. We will then consider different types of integrability breaking perturbations for which collision integrals can be written down explicitly, including atom loss~\cite{10.21468/SciPostPhys.9.4.044} and smooth noise~\cite{PhysRevB.102.161110}. We will also discuss how collision integrals for weak integrability breaking can be formulated perturbatively using Fermi's Golden rule and form factor expansions~\cite{PhysRevB.101.180302,2020arXiv200411030D}. Finally, we will address the limitations of these approaches, and review recent  efforts to approximate collision integrals to address realistic numerical and experimental settings~\cite{2020arXiv200513546L,PhysRevLett.126.090602}. 

\subsection{Diffusive hydrodynamics induced by integrability breaking
\label{sec:diffhydr}
}

GHD describes the dynamics of integrable systems in terms of their quasiparticles. In integrable systems, quasiparticles scatter elastically with phase shifts leading to Wigner time delays, and never decay (there are exceptions with unstable quasiparticles even in the context of integrable systems that we will not consider here, see~\cite{Castro-Alvaredo2020}). Transport properties can be
inferred from the fact that quasiparticles move ballistically and carry some charge (energy, magnetization {\it etc.}): as a result, transport in integrable systems is generically ballistic. We now imagine perturbing such an integrable system with Hamiltonian $\hat{H}_0$ by a small, nonintegrable perturbation $g \hat{U}$ that destroys all but a few conservation laws. Intuitively, the leading effect of the non-integrable perturbation is to thermalize quasiparticle distributions at long times: quasiparticles can scatter into one another, and acquire a finite lifetime. At long times, we expect generic, diffusive hydrodynamics for the residual conserved quantities, characteristic of non-integrable systems. Currents can also overlap with some of the residual conserved quantities, in which case there is ballistic transport even when integrability is broken. This usually requires additional symmetries: prominent examples include Galilean invariant systems where particle transport is ballistic since the particle current is the momentum density, which is conserved; or Lorentz invariant systems where energy transport is ballistic since energy current is a conserved quantity (for a system with emergent Lorentz symmetry, the symmetries of the stress-energy tensor means that the energy current $T_{0i}$ with is also the (conserved) momentum density $T_{i0}$).

Let us first focus on the hydrodynamic equations for the conserved charges of the integrable system $\lbrace \qq_n \rbrace$
\begin{equation} \label{eqGHDcharge}
\partial_t  q_n  + \partial_x j_n = {\cal I}_n \left[ \lbrace q_m \rbrace \right]. 
\end{equation}
Here the left-hand side of those equations corresponds to the ordinary GHD, where $j_n = \langle \jj_n \rangle_{\rm GGE}$ is expressed as a gradient expansion of all the expectation value of the charges $\lbrace q_m \rbrace$. The right-hand ${\cal I}_n$ side encodes the effects of integrability breaking, and will break all but a few conservation laws. We will refer to this term as the collision integral term, for reasons that will become evident below. We will show below that perturbatively, the collision integral is of order ${\cal O}(g^{2})$, leading to thermalization on time scales $ \tau \gg g^{-2}$.
 For now we will not specify the form of this perturbation and treat it in generality. We only enforce that it depends on the charges $ \lbrace q_m \rbrace $ and not on their spatial and temporal derivatives, which is justified in a gradient expansion to zeroth order. In general, the perturbation will also modify the expression of the currents $j_n$, though as we will come back to below, this is a subleading effect perturbatively. We also assume that the system is always in local generalized equilibrium (corresponding to the hydrodynamic regime), with respect to the {\em unperturbed} Hamiltonian. In the following, it will also be convenient to write the GHD equations in the quasiparticle language, as 
\begin{equation} \label{eqGHDqp}
\partial_t  \rho(\lambda)  + \partial_x (v_\lambda^{\rm eff}[\rho] \rho(\lambda)) = {\cal I}_\lambda \left[ \rho \right]. 
\end{equation}
We wrote the right-hand side GHD equations at the Euler scale, $j_\lambda = v_\lambda^{\rm eff}[\rho] \rho(\lambda)$, neglecting diffusive corrections in the integrable limit~\cite{PhysRevLett.121.160603,PhysRevB.98.220303,De_Nardis_2018}, see also Sec.~\ref{sec:diffhydr}. Note that the conservation of the residual charges $q_\alpha$ implies that $\int d\lambda {\cal I}_\lambda h_\alpha(\lambda)=0 $; where $h_\alpha(\lambda)$ is the single particle eigenvalue of the charge $q_\alpha = \int d\lambda \rho(\lambda)  h_\alpha(\lambda)$. In the following, it will be useful to think of eq.~\eqref{eqGHDqp} as a Boltzmann equation for the quasiparticles of interacting integrable systems.

This collision integral will break most conservation laws, but let us assume that it preserves $N$ conserved charges $q_\alpha$, with $\alpha=1,\dots,N$.  In most cases of physical interest, those conserved charges will be energy, spin or particle number, and momentum; while all other charges are broken by the perturbation. We will use Greek letters to denote those residual charges, while $q_{m}$ with $m>N$ are not conserved. We have ${\cal I}_\alpha = 0$ for $\alpha=1,\dots,N$, and we are interested in the long time hydrodynamics of those residual charges. Let us consider linear response on an equilibrium Gibbs state $\rho^{\star} = \frac{1}{Z} {\rm e}^{-\sum_{\alpha=1}^N \beta_\alpha Q_\alpha }$ of those residual conserved charges, with expectation values of the charges denoted by $\lbrace q^\star_m\rbrace$. By assumption, the collision integral evaluated on this Gibbs state vanishes ${\cal I}_n \left[ \lbrace q^\star_m\rbrace \right] =0$ for all $n$, so it describes the steady-state of eq.~\eqref{eqGHDcharge}. Linearizing eq.~\eqref{eqGHDcharge} near this Gibbs state, we find 
\begin{equation} \label{eqLinearizedHydro}
\partial_t \delta q_n  + A_{nm} \partial_x \delta q_m = - \Gamma_{nm} \delta q_m, 
\end{equation}
where repeated indices are implicitly summed over, and $\delta q_n = q_n - q_n^\star$. In the right-hand side of this equation, we evaluated the currents $j_n = A_{nm} \delta q_m$ at the Euler scale (the role of diffusive corrections upon breaking integrability have not been investigated yet), where the matrix  $ A_{nm} = \partial j_n / \partial q_m$ evaluated in the Gibbs state $\rho^{\star}$ is known exactly from GHD at the integrable point $g=0$. In the following, we will neglect corrections in ${\bf A}$ due to the perturbation. This is justified by the fact that while thermodynamic quantities such as the matrix ${\bf A}$ are smoothly affected by the perturbations, with corrections of order ${\cal O}(g)$, the effects of the small perturbation on dynamics at long times are singular. As we will show below, weak integrability breaking perturbations generically lead to a finite conductivity tensor for the residual charges of order ${\cal O}(g^{-2})$, where small corrections to the currents (and to the matrix ${\bf A}$) are subleading for $g \ll 1$. Finally, the matrix ${\bf \Gamma}$ is simply given by $\Gamma_{nm} \equiv - \partial {\cal I}_n / \partial q_m$ evaluated in the Gibbs state $\rho^{\star}$. 

The equations~\eqref{eqLinearizedHydro} already have a simple consequence: if we integrate over space, we find that the total charges decay as $\dot{\delta Q_n} =  - \Gamma_{nm} \delta Q_m$, with $Q_n = \int dx q_n$. The eigenvalues of the matrix ${\bf \Gamma}$ give the decay rates of the quantities ${Q_m}$ that are conserved when $g = 0$. Note that ${\bf \Gamma}$ is positive by definition, since we require ${\rm lim}_{t \to \infty }\delta q_n(t) = 0$ in thermal equilibrium. 
Any residual conserved charge $Q_\alpha$ corresponds to a zero mode (eigenvector with zero eigenvalue) of ${\bf \Gamma}$. 

Our main goal is to determine the dynamics of the residual charges $Q_\alpha$. To get some intuition, let us consider a simplified situation with two charges $q_0$ and $q_1$, with $q_0$ being conserved by the perturbation, while $q_1$ is not. We also assume for simplicity of the argument that without perturbation, the current $j_0 = q_1$ (so $A_{01}=1$). The corresponding hydrodynamic equations read:
\begin{eqnarray}
\partial_t \delta q_0 + \partial_x \delta q_1 &= &0, \notag\\
\partial_t \delta q_1 + \partial_x (A_{10} \delta q_0 + A_{11} \delta q_1) &=& -\Gamma \delta q_1.
\end{eqnarray}
Those equations are admittedly much simpler than~\eqref{eqLinearizedHydro}, but they will illustrate the main physics at play in the general case. Taking a derivative with respect to time of the first equation, and a derivative with respect to $x$ of the second equation, we can easily find an equation for $\delta q_0$ only: $\partial^2_t \delta q_0 + \Gamma \partial_t \delta q_0 =   A_{10} \partial_x^2 \delta q_0  - A_{11} \partial_x \partial_t \delta q_0$. The corresponding dispersion relation is $-\omega^2 - \Gamma i \omega  = - A_{10} k^2 - A_{11} k \omega$, which gives, in the hydrodynamic limit 
\begin{equation}
\omega = - i A_{10} k^2/\Gamma + \dots
\end{equation}
where neglected terms are higher order in $k$. The result of this simple exercise is that at long times and to leading order in the gradient expansion, the dynamics of $q_0$ is {\em diffusive}
\begin{equation} \label{eqDiffusionq0}
\partial_t \delta q_0 =   \frac{A_{10}}{\Gamma} \partial_x^2 \delta q_0  + \dots    
\end{equation}
with a diffusion constant $D = A_{01} A_{10}/\Gamma$ with $A_{01}=1$ in this simple case. This conclusion turns out to be  general, as long as the currents $j_\alpha$ of the residual charges do not couple to the residual charges themselves, and only to the slowly decaying charges $q_n$ with $n>N$. (In most physical situations of interest, such a coupling is forbidden by time reversal symmetry, and can only occur in special cases like particle number for Galilean invariant systems or energy in Lorentz invariant systems.) We emphasize that here diffusion arises
from ``integrating out'' slow but nonconserved degrees of freedom: this mechanism is very different from the diffusive corrections that arise in integrable systems due to the thermal fluctuations of ballistically propagating quasiparticles~\cite{PhysRevLett.121.160603, PhysRevB.98.220303, 10.21468/SciPostPhys.6.4.049}. The approach leading to the diffusion equation~\eqref{eqDiffusionq0} can be extended to the general case~\eqref{eqLinearizedHydro}~\cite{PhysRevB.101.180302}. In the next section, we will instead choose to analyze transport properties from the perspective of the Kubo formula.

\subsection{Kubo formula and conductivity tensor}

In order to analyze the emergence of diffusive hydrodynamics from weak integrability perturbations, we now show that integrability breaking leads to a finite conductivity tensor for the residual charges. From the Kubo formula, the high temperature conductivity tensor is given by 
\begin{equation} \label{eqKubo}
\sigma_{\alpha \beta}(\omega) =   \frac{\beta}{L}  \int_0^\infty \dd t {\rm e}^{i \omega t} \langle \delta J_\alpha (t)  \delta J_\beta (0)  \rangle,
\end{equation}
where $L$ is the system size, $\beta=1/T$ the inverse temperature, and $J_\alpha = \int dx \jj_\alpha(x)$ is the spatially integrated current for the residual charge $q_\alpha$, $\delta J =J - \langle J \rangle$, and where the expectation value is over the Gibbs state $\rho^\star$. We refer the interested reader to, {\it e.g.} the recent review~\cite{RevModPhys.93.025003}, for a detailed discussion of linear response transport and of the Kubo formula.  
In the integrable case, the current  $ J_\alpha $ has hydrodynamic projections on the charges $Q_m$ leading to a finite Drude weight at the Euler level, corresponding to ballistic transport (to leading order)
\begin{equation} \label{eqKubo2}
T \ {\rm Re} \ \sigma_{\alpha \beta}(\omega) =  {\rm Re} \   \int_0^\infty \dd t {\rm e}^{i \omega t} \frac{1}{L} A_{\alpha n } A_{\alpha m }\langle \delta Q_n \delta Q_m  \rangle = \pi \delta (\omega) \left( {\bf A} {\bf C} {\bf A}^T \right)_{\alpha \beta},
\end{equation}
where ${\bf D}={\bf A} {\bf C} {\bf A}^T$ is the Drude weight matrix, with $C_{nm}= \langle \delta Q_n \delta Q_m  \rangle/ L = \int dx \langle \delta \qq_n(x) \delta \qq_m(0)  \rangle  $ is the static susceptibility matrix characterizing equilibrium charge fluctuations. We also emphasize that we have used the fact that the charges are conserved so that $\langle \delta Q_n(t) \delta Q_m(0)  \rangle$ is time independent. 

In the presence of an integrability breaking perturbation, the currents will now decay since the charges $Q_m$ are not conserved anymore. The currents still couple to the slow, quasi-hydrodynamic modes $Q_m$ as before, but we now have a finite d.c. value~\cite{PhysRevB.101.180302,2020arXiv200411030D}
\begin{equation} \label{eqKubo3}
T \sigma_{\alpha \beta} =     \int_0^\infty  \dd t \frac{1}{L} A_{\alpha n } A_{\alpha m }  \left[ {\rm e}^{-{\bf \Gamma} t} \right]_{n k} C_{km} = \left( {\bf A} {\bf \Gamma}^{-1}  {\bf C} \ {\bf A}^T \right)_{\alpha \beta},
\end{equation}
where $\sigma_{\alpha \beta} = \lim_{\omega \to 0} \sigma_{\alpha \beta}(\omega)$, and we have used $\langle \delta Q_n (t) \delta Q_m (0) \rangle/L = \left[ {\rm e}^{-{\bf \Gamma} t} \right]_{n k} C_{km}$. As should be clear from the derivation, the inverse ${\bf \Gamma}^{-1}$ is defined by projecting out zero modes (corresponding to residual conserved charges). The physical picture behind this result is straightforward: eq.~\eqref{eqKubo3} should be thought of as a {\em generalized Drude formula}. Due to the integrability breaking perturbation, charges decay with a rate given by ${\bf \Gamma}$, and the Drude weight~\eqref{eqKubo2} is broadened into Lorentzians in the conductivity $\sigma(\omega)$. Note that we have assumed that the currents of the residual charges $J_\alpha$ do not couple to the residual charges themselves. If they do, as in the case of particle number in Galilean invariant systems for example, those currents will have a conserved part corresponding to a finite Drude weight even upon breaking integrability.
Note that this approach can be applied to compute the decay of arbitrary correlators, using the formalism of hydrodynamic projections~\cite{PhysRevB.101.180302,2020arXiv200411030D}. In general, the matrix ${\bf \Gamma}$ could be ``gapped'' (meaning that after removing zero modes, the slowest modes relax with a finite rate), in which case the corresponding eigenvalues can be interpreted as a relaxation time for the system (and the approach to equilibrium is exponential); or ``gapless'', meaning that arbitrarily long-lived modes survive integrability-breaking. 
Identifying physical cases where ${\bf \Gamma}$ is gapless represents an interesting direction for future works, as it could lead to a divergent conductivity~\eqref{eqKubo3}, corresponding to anomalous (superdiffusive) transport upon breaking integrability (see {\it e.g.}~\cite{2021arXiv210202219D} for a recent example, and Ref.~\cite{2021arXiv210301976B} for a detailed review on anomalous transport in integrable systems in the same volume). Finally, we note that if transport is not ballistic in the integrable limit, the diffusion constants upon breaking integrability can have a non-analytic dependence on the integrability breaking parameter~\cite{PhysRevB.101.180302,PhysRevLett.125.180605}.

\subsection{Atom losses}

So far, we have explored the general consequences of the presence of  collision integral terms in the GHD equations~\eqref{eqGHDcharge}, without specifying their explicit form. In particular, the matrices in Eq.~\eqref{eqKubo3} are known exactly within GHD and can be computed using standard TBA techniques, with the exception of matrix ${\bf \Gamma}$ which depends on the specific form of the collision integral. The rest of this section will be devoted to this question. 

One of the simplest examples is the case of atom losses in the one-dimensional Bose gas with repulsive contact interactions (Lieb-Liniger model). (We refer the reader to the review on the GHD of Bose gases in the same volume for a more extensive discussion of atom losses.) From the perspective of transport and eq.~\eqref{eqKubo3}, atom losses are not the most relevant example since they break all conservation laws, and there is no residual charge to be transported. However, atom losses are usually the main effect of weak coupling to the environment in cold atom experiments, and are interesting in their own right. Here, we follow Ref.~\cite{10.21468/SciPostPhys.9.4.044}, and focus on weak one-body losses (more general $k$-body loss processes were also addressed in this reference). As above, we assume the loss processes are very slow and that the system remains in local equilibrium at all times. (This assumption is valid if the relaxation time of the system is much smaller than the time scale associated with atom losses.) 

In general, an open 1d Bose gas with one-body atom losses can be described by a Lindblad equation for the density matrix $\hat{\rho}$
\begin{equation}
\frac{d {\hat \rho}}{dt} = -i \left[ H_0, {\hat \rho} \right] + g^2 \int dx \left( \psi {\hat \rho} \psi^\dagger - \frac{1}{2} \lbrace \psi^\dagger\psi,  {\hat \rho} \rbrace \right),
\end{equation}
with $H_0 = \int dx \psi^\dagger ( -\frac{\nabla^2}{2m} - \mu) \psi + c \psi^\dagger \psi^\dagger \psi \psi  $ the Lieb-Liniger Hamiltonian. We are using the Lindblad framework here, but this is not necessary, and $g$ can be thought of as the coupling constant between the system and the environment (bath), with which it can exchange particles.  
This equation is not tractable in general, even numerically, and we are after a hydrodynamic description valid in a regime where the atom loss rate is much smaller than the intrinsic relaxation time scale of the system. In other words, we want to identify a ``collision integral'' term in eq.~\eqref{eqGHDqp} that implements one-body losses. (The phrase ``collision integral'' is an abuse of notation here, since there is no collision involved, and simply refers to the right-hand side of  eq.~\eqref{eqGHDqp}.)
This is entirely straightforward in the non-interacting, free boson case ($c=0$), where particle losses lead to a simple decay term 
\begin{equation} 
\partial_t  \rho(\lambda)  + \partial_x (v_\lambda^{\rm eff}[\rho] \rho(\lambda)) = -g^2 \rho(\lambda), 
\end{equation}
since $\rho(\lambda)$ is simply the density of bosons with momentum $\lambda$ in that case. However, in the presence of interactions $c \neq 0$, the quasiparticle excitations become distinct from the physical bosons, and the collision integral in this case is a lot more involved. As we will see more explicitly below, the collision integral involves matrix elements of the integrability perturbation, here $\psi$, between different generalized equilibrium states (characterized by different quasiparticle distributions $\rho(\lambda)$), and summing over states. Even in this simple case of particle loss, this turns out to be very hard to do explicitly without any approximation, and was implemented numerically in Ref.~\cite{10.21468/SciPostPhys.9.4.044}, in a way that does not appear practical for GHD applications at the moment. This already highlights one of the key difficulties in computing collision integrals analytically, even in a perturbative regime. We will come back to this point later. 

In the hard-core (Tonks–Girardeau) limit $c\to \infty$, the Lieb-Liniger model maps to non-interacting fermions. However, the collision integral term is still highly non-trivial because of the Jordan-Wigner string relating fermions and bosons. A closed form formula was derived in Ref.~\cite{10.21468/SciPostPhys.9.4.044}, and ends up fairly intricate, even though the underlying model can be mapped onto non-interacting fermionic particles. 

Away from these special limits, the variation $d \rho (\lambda)/d t$ can be computed for a finite system size $\ell$ as follows, as shown in Ref.~\cite{10.21468/SciPostPhys.9.4.044}. Say the system is in a Bethe state $\Ket{\lbrace \lambda_i \rbrace }$ sampled with a probability $p_\rho(\lbrace \lambda_i \rbrace)$ to represent the rapidity distribution $\rho(\lambda)$. In a time step $dt$, the probability that the system is found in a different state $\Ket{\lbrace \lambda'_i \rbrace }$ due to the perturbation is $L g^2 dt \left| \langle \lbrace \lambda'_i \rbrace  \left| \psi \right| \lbrace \lambda_i \rbrace  \rangle \right|^2  $    in which case the new rapidity distribution is $\frac{1}{\ell} \sum_i \delta(\lambda - \lambda'_i)$. On the other hand, the probability for the system to remain in the state $\Ket{\lbrace \lambda_i \rbrace }$ corresponding to rapidity distribution $\frac{1}{\ell} \sum_i \delta(\lambda - \lambda_i)$ is $1-\ell g^2 dt \sum_{\lbrace \lambda'_i \rbrace} \left| \langle \lbrace \lambda'_i \rbrace  \left| \psi \right| \lbrace \lambda_i \rbrace  \rangle \right|^2 = 1 - \ell g^2 dt \langle \psi^\dagger \psi \rangle$. As a result, we have~\cite{10.21468/SciPostPhys.9.4.044}
\begin{equation}
{\cal I}_\lambda[\rho] = g^2 \sum_{\lbrace \lambda_i \rbrace} p_\rho(\lbrace \lambda_i \rbrace) \sum_{\lbrace \lambda'_i \rbrace} \left| \langle \lbrace \lambda'_i \rbrace  \left| \psi \right| \lbrace \lambda_i \rbrace  \rangle \right|^2 \left( \sum_i \delta(\lambda - \lambda'_i) - \sum_i \delta(\lambda - \lambda_i) \right). \label{eqParticlelossIntegral}
\end{equation}
For large enough systems, the sum over states $\lbrace \lambda_i \rbrace$ can in principle be dropped in favor of a single representative eigenstate, but this is not practical numerically for accessible sizes. 
Using recent formulas~\cite{PiroliCalabrese_2015,Pozsgay_2011P11017} for the form factors $\langle \lbrace \lambda'_i \rbrace  \left| \psi \right| \lbrace \lambda_i \rbrace  \rangle $, Ref.~\cite{10.21468/SciPostPhys.9.4.044} managed to sample this double sum numerically to evaluate ${\cal I}_\lambda$. This involves a number of numerical steps including numerically solving the Bethe equations in finite size, evaluating matrix elements efficiently and properly sampling those sums. More explicit expressions were also obtained for energy loss~\cite{hutsalyuk2021integrability}, opening the door to more efficient simulations of GHD equations with atom losses in the near future. 

While this approach could in principle be generalized to other perturbations, it remains very demanding numerically and requires working with finite systems. In the next section, we will discuss how to write down collision integrals directly in the hydrodynamic limit using the matrix elements (thermal form factors) of the perturbation, in different cases involving low momentum transfer. 

Closed and manageable expressions can be obtained for strong two-body losses \cite{rossini2020strong}. In this limit, double-occupied sites decay on a very short time-scale imposing a hard-core constraint even for finite values of the interactions and thus mapping the system into free fermions.

\subsection{Fermi Golden Rule and thermal form factors}

In principle, collision integrals can be computed perturbatively using Fermi Golden Rule (FGR) (see Ref.~\cite{PhysRevX.9.021027} in the context of interacting systems), as in standard Boltzmann equations for weakly interacting particles.
As we will see below, in practice this requires summing over states and knowing the matrix elements (form factors) of the perturbation between a given (generalized) equilibrium state $\rho$ and arbitrary excited states. Both of these steps are complicated in general, but we will see below how for certain types of perturbations that restrict momentum transfer, only a handful of particle-hole excitations contribute, and the matrix elements take a simple ``hydrodynamic'' form in terms of GHD data. 

For a given perturbation, the general structure of the collision integral in the Boltzmann equation~\eqref{eqGHDqp} can be largely extrapolated from the well-known non-interacting case, thinking of this equation as a master equation in rapidity space. In particular, the collision integral will include positive contributions (``in-scattering'' terms) corresponding to processes that increase $\rho(\lambda)$ and negative contributions (``out-scattering'' terms) corresponding to a sum over all processes contributing to making $\rho(\lambda)$ decay~\cite{PhysRevB.101.180302}. The rates at which those processes occur can be inferred from FGR, and are given by the square of the matrix elements times density of state factors, with delta functions ensuring residual conservation laws such as energy or momentum. In particular, all those rates and thus the collision integral are of order ${\cal O}(g^2)$ with $g$ the strength of the perturbation. All we will see below, this intuition from weakly-interacting particles is largely correct, although there are additional effects specific to interactions. In particular, because of the structure of the Bethe equations, a scattering process involving rapidities different from $\lambda$ can still affect $\rho(\lambda)$ through ``backflow'' effects, leading to additional contributions to the collision integral~\cite{2020arXiv200411030D}. 

In order to derive the form of the collision integral, we start from the evolution of the expectation value of the conserved charges of the integrable model in question, using standard second order perturbation theory (see e.g.~\cite{2020arXiv200411030D})
\begin{equation} 
\partial_t \langle q_k \rangle  = g^2 \int_{-\infty}^{\infty} ds \langle [{\rm e}^{i \hat{H}_0 s} \hat{U} {\rm e}^{-i \hat{H}_0 s}, \hat{Q}_k ] \hat{u} \rangle_c. \label{eqPerturbationTheory}
\end{equation}
Recall that $\hat{H} = \hat{H}_0 + g \hat{U}$ with $\hat{H}_0$ the integrable Hamiltonian, and $\hat{U} = \int dx \hat{u}(x)$. We consider spatially homogeneous situations for now, and will reintroduce spatial gradients later on. 
In this expression~\eqref{eqPerturbationTheory}, we already considered the limit of long times and small perturbations, with $t g^2$ being held fixed. The connected correlation function on the right-hand side is evaluated in an instantaneous generalized equilibrium state, with corresponding quasiparticle density $\rho(\lambda)$. In the following, we will drop the expectation value $q_k = \langle q_k \rangle$, and indicate quantum operators with hats.  
To evaluate this correlator, we insert a resolution of the identity
\begin{equation} 
\partial_t  q_k   = g^2 \sum_n \int_{-\infty}^{\infty} ds \langle \rho | [{\rm e}^{i \hat{H}_0 s} \hat{U} {\rm e}^{-i \hat{H}_0 s}, \hat{Q}_k ] | n \rangle \langle n| {\hat u} | \rho \rangle, 
\end{equation}
where the states $| n \rangle$ diagonalize the (unperturbed) Hamiltonian $\hat{H}_0$ and conserved charges $\hat{Q}_k$. Computing the integral over $s$, we find
\begin{equation} \label{eqFGRcharge}
\partial_t  q_k   =4 \pi^2 g^2 \sum_n \delta (\Delta \varepsilon) \delta (\Delta p) \Delta q_k 
| \langle n| \hat{u} | \rho \rangle |^2,
\end{equation}
where $\Delta q_k $ is the difference of the eigenvalue of $\hat{Q}_k$ on the state $| n \rangle$ compared to the reference state $| \rho \rangle$;
and similarly for $\Delta \epsilon$ where we recall that $\varepsilon$ is the energy eigenvalue, and $\Delta p$ with $p$ the momentum. As expected, the delta functions restrict the sum over processes respecting energy and momentum conservation, since energy is a residual conserved quantity here as the perturbation is time-independent. Equation~\eqref{eqFGRcharge} has a simple FGR form for the change of conserved charges, summing over allowed processes with FGR rates given by the square of the matrix elements of the perturbations.

At this point, it is worth pointing out that this formula can be readily generalized to time-dependent, noisy perturbations
$\hat{H}(t) =\hat{H}_0 + g \int dx \eta(x,t) {\hat v}(x) $ with noise correlations given by 
\begin{equation}
\label{eq:noisecorr}
\langle \eta(x,t) \eta(x',t') \rangle = G(x-x') F(t-t')\;.
\end{equation}
In this case, energy and momentum are not conserved, and eq.~\eqref{eqFGRcharge} becomes
\begin{equation} \label{eqFGRchargeNoise}
\partial_t  q_k   = g^2 \sum_n  \tilde{G} (\Delta p) \tilde{F} (\Delta \varepsilon) \Delta q_k 
\left| \langle n| \hat{u} | \rho \rangle \right|^2,
\end{equation}
where $\tilde{G} (p) = \int dx {\rm e}^{-i p x} G(x)$ is the Fourier transform of the noise spatial correlations, and similarly, $\tilde{F} (\varepsilon) = \int dt {\rm e}^{-i \varepsilon t} F(t)$. 

Equations~\eqref{eqFGRcharge} and~\eqref{eqFGRchargeNoise} have an appealing form, and it is possible to write down the sum over accessible states $\sum_n$ in terms of general particle-hole excitations over the background state  $| \rho \rangle$~\cite{2020arXiv200411030D}.
However, we now face two major hurdles: \emph{i)} The matrix elements (form factors) $\left| \langle n| \hat{u} | \rho \rangle \right|$ of the perturbation are unfortunately not known in general,
and \emph{ii)} Even if we did know the matrix elements, the sum over accessible states would likely be impossible to carry out analytically. To proceed, we need the matrix elements to fall within GHD, restricting the type of perturbations one can deal with. We thus consider perturbations that cause only small momentum transfers, and have a chance of being ``hydrodynamic''. Examples include long-range interaction potentials (where the long-range nature of the interaction can only lead to small momentum processes), or smoothly varying (correlated) noise~\cite{PhysRevB.101.180302,PhysRevB.102.161110} that we will discuss in more detail below. In those cases, the momentum transfer $\Delta p$ is small, and the sum over accessible states can be truncated to take into account single or two-particle hole excitations only. The corresponding form factors are known exactly, and fall naturally into the framework of GHD (We refer the reader to the review on thermal form factors in this special issue). 

In the rest of this subsection, we will assume that the perturbation $\hat{u}$ is such that we can restrict the sum over accessible states to single particle-hole (1ph) excitations $\lambda \to \lambda^\prime$. (A concrete example that falls into this category will be discussed below.) Considering only 1ph excitations, we have
\begin{equation} 
\label{eq:noise1ph}
\partial_t  q_k   = g^2 \int d \lambda \int d \lambda^\prime \rho(\lambda) \rho^h (\lambda^\prime) \tilde{G}(\Delta p) \tilde{F}(\Delta \varepsilon) \Delta q_k \left| \langle \rho| \hat{u} | \rho; \lambda \to \lambda^\prime \rangle \right|^2, 
\end{equation}
where $\rho^h(\lambda^\prime)$ is the density of holes at rapidity $\lambda^\prime$. To convert this expression for the decay of the charges to a collision integral for $\rho(\lambda)$, the last step is to write down $\Delta q_k$ explicitly. We proceed in two steps, by first ignoring backflow effects, in which case we simply have $\Delta q_k = h_k (\lambda^\prime) - h_k (\lambda)$. Using the symmetry $\lambda \leftrightarrow \lambda^\prime$ of the form factor~\cite{De_Nardis_2018}, we have $\partial_t  q_k  =  g^2 \int d \lambda h_k(\lambda) \int d \lambda^\prime \left( \rho(\lambda^\prime) \rho^h(\lambda) - \rho(\lambda) \rho^h(\lambda^\prime) \right) {\tilde G}(\Delta p) \tilde{F}(\Delta \varepsilon)  \left| \langle \rho| \hat{u} | \rho; \lambda \to \lambda^\prime \rangle \right|^2 $. Matching this expression to 
$\partial_t  q_k   = \int d \lambda h_k(\lambda) \partial_t \rho(\lambda)$, we identify the collision integral corresponding to in and out 1ph scattering processes 
\begin{align} \label{eqInOutCollisionIntegralNoise}
{\cal I}_\lambda^{\rm in, out}  = & g^2 \int d \lambda^\prime   {\tilde G}(\Delta p) \tilde{F}(\Delta \varepsilon)   \left| \langle \rho| u | \rho; \lambda \to \lambda^\prime \rangle \right|^2 \notag \\
& \times  \rho^{t}(\lambda) \rho^{t}(\lambda^\prime) \left[ n(\lambda^\prime) (1 - n(\lambda)) - n(\lambda) (1 - n(\lambda^\prime))  \right],
\end{align}
where we have introduced the total density of states $\rho^t = \rho + \rho^h$. The physical interpretation of this equation is straightforward~\cite{2020arXiv200411030D}: the changes in $\rho(\lambda)$ come from 1ph excitations $\lambda^\prime \to \lambda$, minus all out-scattering events $\lambda \to \lambda^\prime$. The corresponding rates follow from FGR, and are given by the square of the matrix elements $g^2  {\tilde G}(\Delta p) \tilde{F}(\Delta \varepsilon)   \left| \langle \rho| \hat{q}_0 | \rho; \lambda \to \lambda^\prime \rangle \right|^2 $ times density of states factors.

In the presence of interactions, the expression of $\Delta q_k$ is more complicated, and is given by $\Delta q_k = \left[(1+\partial({\bf F n})) h_k \right] (\lambda^\prime) - \left[(1+\partial({\bf F n})) h_k \right] (\lambda)$, where $F(\lambda, \lambda^\prime)$ is the so-called backflow function representing the effect of adding an excitation to the interacting background $| \rho \rangle$. Here we are using an operator notation introduced in \eqref{eq:opnot} where $\bf{F}$ is an operator (matrix) in rapidity space. Using this compact notation, we have $\left[(1+\partial({\bf F n})) h_k \right] (\lambda) = h_k(\lambda) + \int d \lambda^\prime \partial_{\lambda^\prime} \left( F(\lambda,\lambda^\prime) n(\lambda^\prime) \right) h_k(\lambda^\prime) $. The backflow function satisfies the integral equation $2 \pi F(\lambda,\lambda') =  \Theta(\lambda -\lambda') + \int d \alpha \partial_\lambda \Theta(\lambda - \alpha) F(\alpha,\lambda')$.
Repeating the previous steps with this modified expression for $\Delta q_k$, we find $\partial_t \langle q_k \rangle = \int d \lambda  \left[(1+\partial({\bf F} {\bf n})) h_k \right] (\lambda) {\cal I}_\lambda^{\rm in, out}$, so the total collision integral including backflow processes reads 
\begin{equation} \label{eqFinalCollisionNoise}
{\cal I} = (1+\partial({\bf n} {\bf F}^{T} )) {\cal I}^{\rm in, out},
\end{equation}
with ${\bf F}^{T} $ the transpose of ${\bf F}$, or more explicitly, $I_\lambda = \int d\lambda' \left[1 + \partial({\bf n F}^T) \right]_{\lambda, \lambda^\prime} {\cal I}^{\rm in, out}_{\lambda^\prime}$. Intuitively, backflow effects take into account 1ph excitations $\lambda^\prime \to \lambda^{\prime \prime}$ that affect $\rho(\lambda)$ indirectly through the Bethe equation.  Equation~\eqref{eqFinalCollisionNoise} is the final expression of our collision integral restricting to 1ph excitations. The matrix element $\left| \langle \rho| u | \rho; \lambda \to \lambda^\prime \rangle \right|^2$ remains to be determined, and we will discuss a case where it can be explicitly expressed in terms of GHD data in the next section.  

It is possible to write down collision integrals much more generally as a formal infinite series over particle-hole excitations, involving (unknown) form factors~\cite{2020arXiv200411030D}. The general schematic form of the collision integral is as follows (for say, a static, translation-invariant perturbation): it will now involve a sum over particle-hole excitations 
\begin{align} \label{eqInOutCollisionIntegralMultiple}
{\cal I}_\lambda  &= \sum_{\rm p.h. \ processes} g^2 \int \prod_{ij} d \alpha_i d \beta_j     (2 \pi)^2 \delta(\Delta \varepsilon ) \delta(\Delta p ) \left| \langle \rho| \hat{u} | \rho; {\vec \alpha} \to {\vec \beta} \rangle \right|^2 K(\lambda, {\vec \beta}) \notag \\
& \times \prod_{ij} \rho^{t}(\alpha_i) \rho^{t}(\beta_j) \left( \prod_i n(\alpha_i) \prod_j (1 - n(\beta_j)) - \prod_j n(\beta_j) \prod_i (1 - n(\alpha_i))  \right), 
\end{align}
where the function $K$ encodes backflow effects similar to those discussed above~\cite{2020arXiv200411030D}. We have $K(\lambda, {\vec \beta}) = \sum_j \delta(\beta_j - \lambda)$ in the case where backflow effects are ignored: one of the rapidities $\beta_j$'s must coincide with $\lambda$ so that the collision integral corresponds to a sum over in- and out-scattering events. 
The scattering can happen in various permutations, which must be summed over, and the delta functions ensure that energy and momentum are conserved. As we will argue below, such a general expansion is likely to be practical only for perturbations allowing for small momentum transfers, restricting dramatically its utility in general. It will be interesting to investigate whether future progress towards understanding  higher-order form factors can allow one to tackle more generic integrability breaking perturbations.

\subsection{Smoothly varying noise}

\label{secNoise}

As a concrete example of integrability perturbation that can be treated analytically within GHD, we consider smooth (spatially correlated) noise coupled to the charge density $\hat{q}_0$~\cite{PhysRevB.102.161110}. We will focus on noise that is uncorrelated in time, so that $F(t) =\delta(t)$ and $\tilde{F}=1$ in eq.~\eqref{eqFGRchargeNoise}. Let us further assume that the noise is very smooth, so that $\tilde{G}$ is very narrow in momentum space (close to a delta function). As has been argued in Refs.~\cite{De_Nardis_2018,10.21468/SciPostPhys.6.4.049}, this limit of small momentum transfer controls the particle-hole expansion, so that the 1ph expression~\eqref{eqFinalCollisionNoise} becomes accurate, with negligible higher-order terms. Moreover, in this small momentum transfer limit, the matrix elements also reduce to the dressed charges
$\lim_{\lambda^\prime \to \lambda}\langle \rho|\hat{q}_{\barj}| \rho; \lambda \to \lambda^\prime \rangle = q^{\rm dr}_{\barj}(\lambda)$~\cite{De_Nardis_2018,10.21468/SciPostPhys.6.4.049}. Those two dramatic simplifications imply that for very smooth noise, the collision integrals are precisely given by eqs.~\eqref{eq:noise1ph},~\eqref{eqFinalCollisionNoise} , with an explicit form of the matrix element of the perturbation in terms of the dressed charge. 

\begin{figure}[ht]
         \centering
           \includegraphics[width=0.45\textwidth]{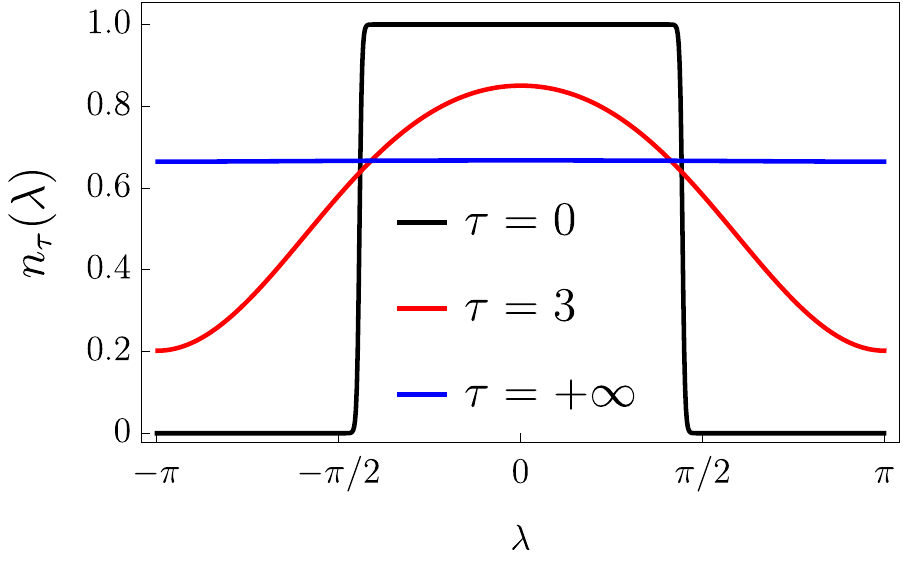}                        \includegraphics[width=0.45\textwidth]{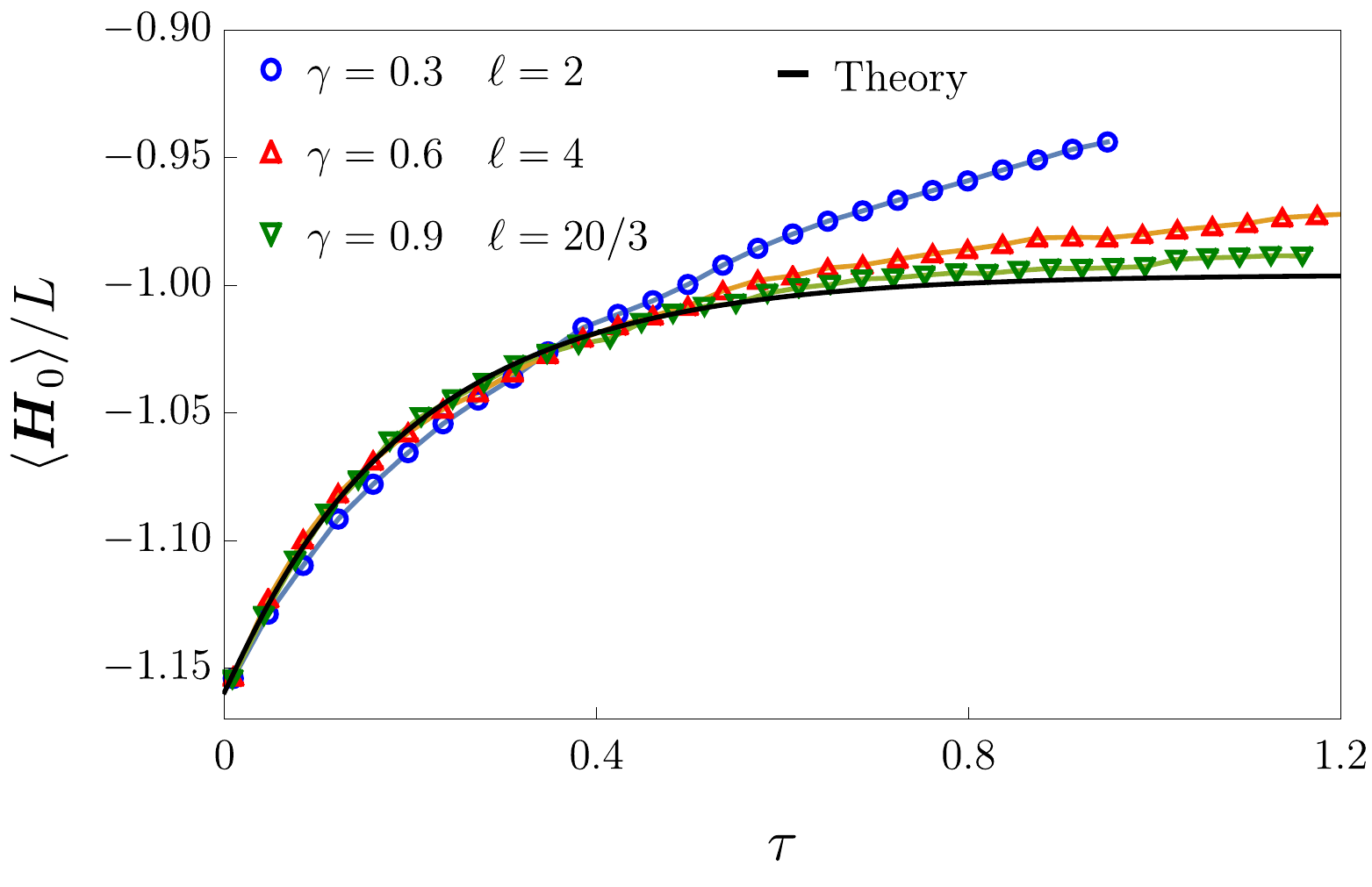}
    \caption{ Behavior of the XXZ spin chain at $\Delta = \cosh(3/2)$ initially prepared in the ground state at magnetization $1/10$ and perturbed by the presence of noise coupled to the local magnetization $H = H_{\rm XXZ} + \sum_i \eta_i(t) S_j^z$. The noise correlation is chosen as in Eq.~\eqref{eq:noisetay} and time has been rescaled as $\tau = g^2 \kappa t/ \ell$. Left: Behavior of the filling function $n(\lambda)$ under the diffusive dynamics of Eq.~\eqref{eq:noisediff}. The initial Fermi sea (groundstate) melts because of noise. At infinite time, the stationary state is reached where the filling is a constant function in the rapidity space. Right (from Ref. \cite{PhysRevB.102.161110}): Behavior of the energy density under the same dynamics compared with numerical simulations (see \cite{PhysRevB.102.161110}).}
    \label{noisyXXZ}
\end{figure}

We can further use the form of the function $\tilde{G}(\Delta p)$ to simplify the double integral in $\lambda$ and $\lambda'$. To do this more systematically, we assume that the noise is correlated on a length scale $\ell$, by taking a scaling form for its spatial correlation as $G(x) = \ell \mathfrak{g}(x/\ell)$ where $\mathfrak{g}(y)$ is a smooth and even function which decays at infinity. Expanding at large $\ell$ and taking the Fourier transform, we obtain
\begin{equation}
\label{eq:noisetay}
    \tilde G(\Delta p) = 2\pi \ell \mathfrak{g}(0) \delta (\Delta p) + \frac{2\pi \kappa}{\ell} \delta''(\Delta p) + {\cal O}(\ell^{-3}),
\end{equation}
where we set $\kappa = - \mathfrak{g}''(0)/2>0$. We can inject this expansion in Eq.~\eqref{eqInOutCollisionIntegralNoise} and perform the integral over $\lambda'$. As expected the leading order ${\cal O}(\ell)$ in \eqref{eq:noisetay} does not contribute because it corresponds to coupling the system to a global conserved quantity. On the contrary, at the order ${\cal O}(\ell^{-1})$, we obtain after integrating over the $\delta''(\Delta p)$
\begin{equation}
\partial_t \rho(\lambda) = \frac{g^2 \kappa}{2 \pi \ell} \partial_\lambda\left( (1 + \frac{\bf{n} \boldsymbol{\varphi}}{2\pi} )^{-1}  q^{\rm dr}_0(\lambda)^2  \frac{1}{p'}\partial_\lambda n \right).
\end{equation}
and the vector notation introduced in \eqref{eq:opnot} as been used. 
Further intuition can be gained by rewriting this equation only in terms of the filling function $n(\lambda) = \rho(\lambda) / \rho_t(\lambda) = \rho(\lambda) / (2\pi p'(\lambda))$, using that
    $\partial_t \rho(\lambda) = (1 + \boldsymbol{n} \boldsymbol{\varphi})^{-1} [\rho_t \partial_t n(\lambda)]$.
This leads to 
\begin{equation}
\label{eq:noisen}
    \partial_t n(\lambda) = \frac{g^2 \kappa}{ p' \ell} \left[ \partial_\lambda \Bigl((q^{\rm dr}_0)^2  \frac{1}{p'}\partial_\lambda n \Bigr) - \partial_\lambda n \; \boldsymbol{\varphi}^{\rm dr} \Bigl(\frac{1}{p'}\partial_\lambda n\Bigr)\right].
\end{equation}
Then, we reparameterize the rapidity space in terms of the dressed momentum. To do so, 
we define implicitly the filling $\mathfrak{n}(p)$ as a function of the quasiparticle momentum $p$ via 
\begin{equation}
    \mathfrak{n}(p(\lambda)) = n(\lambda) \;, \qquad p(\lambda) = 2\pi \int_{\lambda_0}^\lambda d\lambda' \rho_t(\lambda') \;.
\end{equation}
In the following we assume that the boundary value $\lambda_0$ can be chosen such that $n(\lambda = \lambda_0)$ remains constant in time. This can be done for instance by choosing $\lambda_0 \to -\infty$ for non-compact rapidity space. We can thus replace $\partial_p  = (p')^{-1}\partial_\lambda$. Instead, for the time derivative one has to account for the time variation of the quasimomentum due to dressing
\begin{equation}
\label{eq:nlt}
\partial_t n = \partial_t \mathfrak{n} + \partial_p \mathfrak{n} \; \partial_t p    \,.
\end{equation}
Explicit calculations show that the second term in \eqref{eq:nlt} cancels exactly the second term in \eqref{eq:noisen}. One thus arrives at
\begin{equation}
    \label{eq:noisediff}
\partial_t \mathfrak{n} = \frac{g^2 \kappa}{\ell} \partial_p [(q_0^{\rm dr})^2 \partial_p \mathfrak{n}],
\end{equation}
which has the form of a diffusion equation in momentum space, generalizing a well-known effect of noise in non-interacting systems to strongly interacting settings.  Note that because of the dressing operation, the diffusion constant $\propto (q_0^{\rm dr})^2$ depends on the state itself $\mathfrak{n}(p)$, making the equation highly nonlinear. 
An example of the solution of this equation is shown in Fig.~\ref{noisyXXZ} (left panel).

\subsection{Approximate expressions}

While the perturbative approach outlined in the previous section combined with the exact expressions of form factors provides an appealing, systematic way to tackle integrability breaking, we emphasize that its scope is extremely limited for now. In fact, the only tractable case to date is that of smoothly varying noise. Other possible examples would involve long-range interactions which also restrict momentum exchange. On the other hand, most integrability-breaking perturbations of interest involve many particle-hole  processes with possibly large momenta exchange: for example, Umklapp scattering involves large momentum transfer, and is thus unlikely to be captured by a long-wavelength theory such as GHD which involves only 1ph and 2ph processes~\cite{De_Nardis_2018}. While it is conceivable that future works on form factors will allow one to address more general scattering processes, we note that this obstruction appears to be fundamental rather than merely technical. Generic integrability-breaking processes involve large momentum transfer, and are beyond hydrodynamics by nature. In the absence of  simple GHD expressions for the matrix elements, evaluating collision integrals is at the moment an intractable task, and would require evaluating the sum over accessible states and evaluating the corresponding form factors by  solving Bethe equations on a lattice and/or for a finite number of particles, as in Eq.~\ref{eqParticlelossIntegral}.

\begin{figure}
         \centering
           \includegraphics[width=0.6\textwidth]{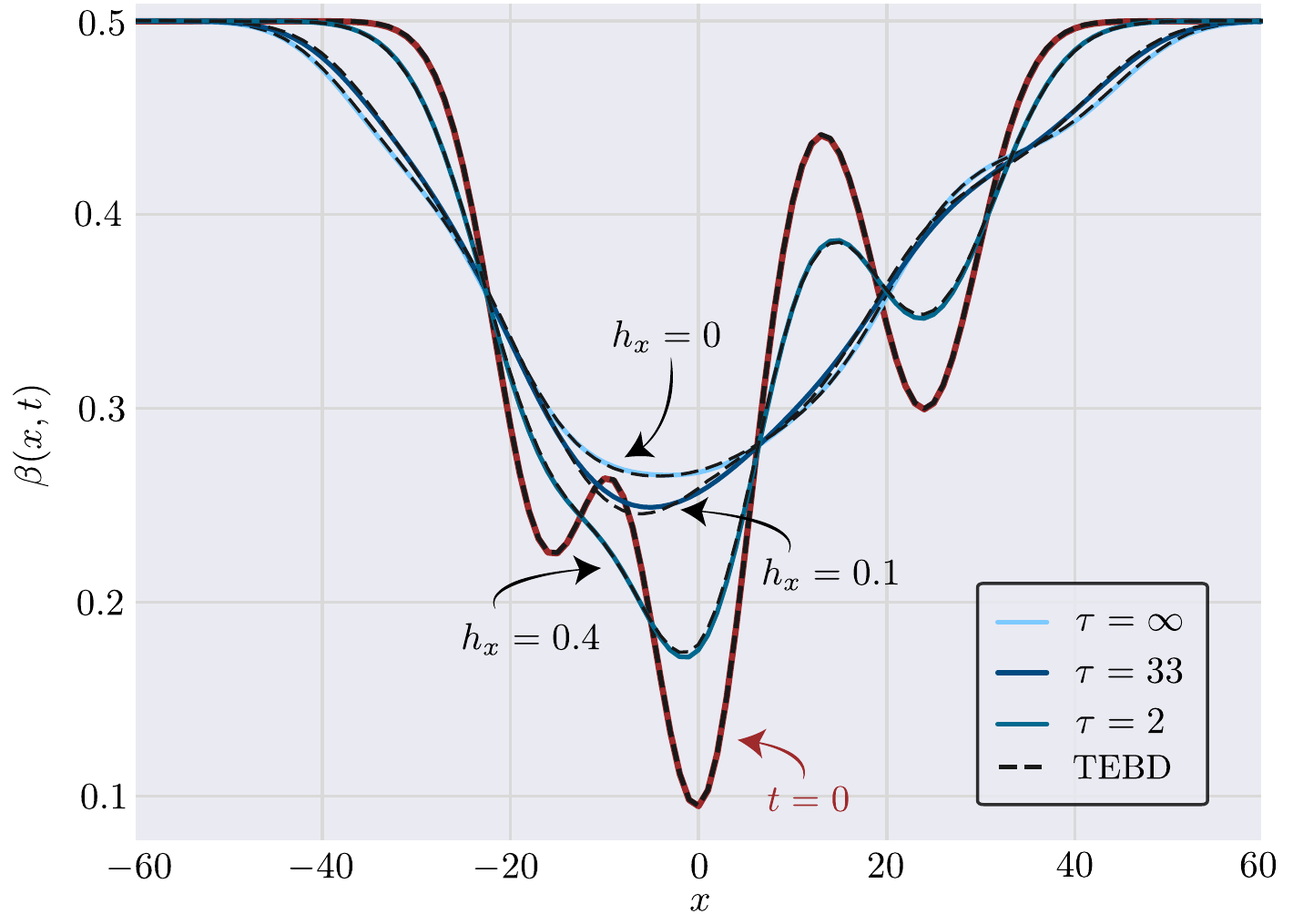}
 
    \caption{ Local temperature profiles in an XXZ spin chain with staggered magnetic field $h_x \sum_i (-1)^i S_i^x$ breaking integrability, at fixed time $t=20$ for various values of $h_x$. The initial state is a locally thermal state with temperature profile plotted in red. The numerical data from time-evolving block decimation (TEBD) agrees very well with the prediction from the GHD equations with generalized RTA collision integral~\eqref{eqGRTA}. The value of the phenomenological relaxation time $\tau(h_x)$ was fitted numerically for a different initial state, and obeys the Fermi Golden Rule scaling $\tau \sim h_x^{-2}$. 
    Figure reproduced from Ref.~\cite{2020arXiv200513546L}.}
    \label{fig:GRTA}
\end{figure}

In this context, it is crucial to develop simple but accurate approximations to collision integrals. Of course, this idea is not new, and collision integrals are routinely approximated using the so-called relaxation time approximation (RTA) to describe scattering processes in Fermi liquids for example. In most cases, crude approximations to collision integrals are enough to describe experiments. The development of such approximations for interacting integrable systems remains in its infancy. 
Here, we briefly summarize two recent attempts:
\begin{itemize}
    \item {\it Non-interacting expressions:} One approach considered in Ref.~\cite{PhysRevLett.126.090602} is to ignore the effects of interactions when writing collision integrals, and to use similar forms as in the non-interacting literature. In particular, Ref.~\cite{PhysRevLett.126.090602} studied dimensional crossovers in Bose gases by writing down collision integrals inspired from the non-interacting case in particular, approximating momentum with rapidity, and quasiparticles with physical bosons. While such an approximation is not expected to be valid in the presence of strong interactions (in particular, it would be hard to enforce residual conservation laws using this approach), Ref.~\cite{PhysRevLett.126.090602} showed that the thermalization time extracted from such collision integrals was compatible with experimental data. 

    \item {\it Generalized RTA:} Another approach proposed in Ref.~\cite{2020arXiv200513546L} is to generalize the RTA to integrable systems, by approximating the collision integral by a term that enforces {\em local} thermalization to a Gibbs ensemble on a single time scale $\tau$. This is expected to capture the main physics of integrability breaking, and is expected to be a good approximation at long times if the spectrum of the matrix ${\bf \Gamma}$ in eq.~\eqref{eqLinearizedHydro} is gapped, in which case one can naturally identify the relaxation time $\tau$ with the inverse of the gap. Within this generalized RTA, one writes
\begin{equation} \label{eqGRTA}
\partial_t  \rho  + \partial_x (v^{\rm eff}[\rho] \rho) = -(\rho - \rho^{\rm Gibbs}[\rho])/\tau, 
\end{equation}
where $\rho^{\rm Gibbs}[\rho]$ is the root density corresponding to a Gibbs ensemble of the residual conserved quantities (say energy), with the same value of those residual conserved quantities as in the state $\rho$. This enforces local thermalization, and also preserves the residual conservation laws contrary to the standard RTA which is not conserving. The downside of this approach (besides the crude approximation of the collision integral) is that it introduces $\tau$ as a free, phenomenological parameter. 
Nevertheless, in the limit of short $\tau$, this type of method was used to investigate transport in the XXZ spin chain in the presence of a localized non-integrable defect~\cite{Biella2019}.
More recently, Ref.~\cite{2020arXiv200513546L} showed that such an approximation captured remarkably well (with errors of order 1 or 2 percents) heat transport even for far-from-equilibrium initial states in strongly chaotic spin chains, relatively far from integrable limits. The relaxation time was also found to scale as $\tau \sim g^{-2} $, as expected from Fermi Golden Rule considerations. (See discussion above.) 

\end{itemize}

More work in this direction is clearly needed to establish more accurate approximations, and to estimate relaxation times analytically (as in the case of Fermi liquids for example). At the same time, progress is needed to evaluate form factors within GHD, and to tackle a broader variety of experimentally-relevant perturbations.

\section{Hidden non-adiabaticity and bound states recombination }
\label{sec_hidd_nonadiab}

So far, we have seen how the hydrodynamic approach can describe weak integrability-breaking terms, smoothly modifying the local root density due to ballistic expansion and force terms, with the further possibility of including diffusive corrections and collisional integrals. All these scenarios have in common a fixed, or at least smoothly varying, integrable model of reference, characterized by certain excitations' species and scattering data.
Indeed, one of the cornerstones of integrability is the stability of the underlying excitations: from the GHD point of view presented so far, these excitations can be transported or accelerated, either in a deterministic way (Eulerian GHD) or stochastically (diffusion and noise).
However, integrable models are very sensible to the interactions and apparently smooth inhomogeneities can bring the model within a new phase, where some excitations can cease to exist and others appear in the spectrum. Physically, one faces the challenge of incorporating particle-recombination in the theory.
In this scenario, the applicability of GHD seems at stake, due to the very non-integrable nature of unstable particles: however, it can be still apllied under certain conditions.
The theory behind integrable models with such drastic interaction changes has not been addressed in complete generality yet, but two examples have been considered in Ref. \cite{PhysRevLett.122.240606} and Ref. \cite{koch2020adiabatic}.
In this section, we review the content of these two papers. 

For the sake of clarity, it is useful to focus on the specific models studied in Refs. \cite{PhysRevLett.122.240606,koch2020adiabatic}, but the methods are expected to be of wider applicability. At the end of this section, we will comment on possible future generalizations beyond the models here discussed.
Here after, we revert the chronological order among Refs. \cite{PhysRevLett.122.240606} and \cite{koch2020adiabatic} and start discussing the second, which is physically more intuitive and needs less technicalities.

\subsection{Bound state recombination in the interacting Bose gas}
\label{eq_sec_attractive_bose_gas}

Following Ref. \cite{koch2020adiabatic}, we are now interested in the 1d interacting Bose Gas, which we have already extensively discussed Sections \ref{sec_firstorder_lackth} and \ref{sec_force_second_order}.
In Section \ref{sec_firstorder_lackth} we focused on the effects of the trap while keeping the interaction constant $c>0$ in the Hamiltonian \eqref{eq_LL_H}, whereas here we consider interaction changes. In particular, we are willing to study smooth sign changes in the interaction.
It is worth to be mentioned that such a protocol is within the reach of the state-of-art cold atom experiments \cite{Haller1224,Kao296}, where the effective interaction in the 1d system can be arbitrarily tuned.

\begin{figure}
        \begin{minipage}{.4\textwidth}
         \centering
           \includegraphics[width=1\textwidth]{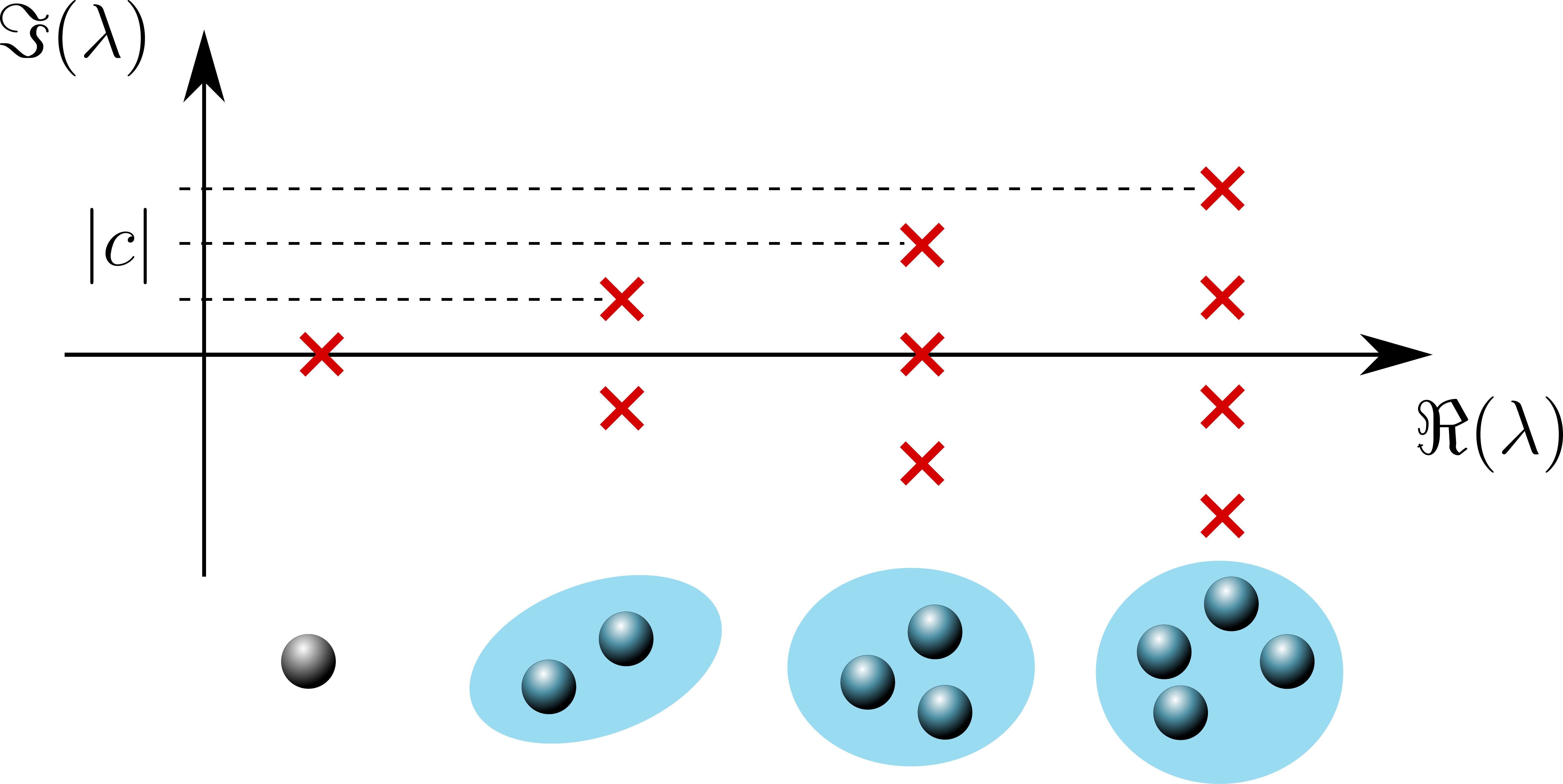}
    \end{minipage}%
    \begin{minipage}{0.6\textwidth}
        \centering
       \ \\
         \includegraphics[width=0.8\textwidth]{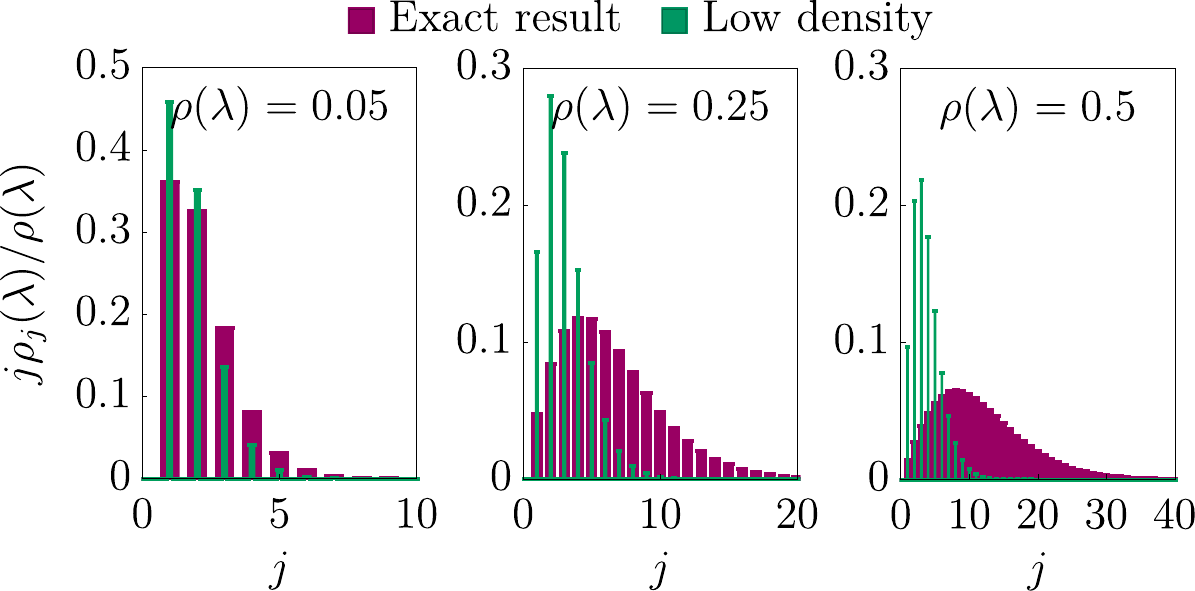}
    \end{minipage}
    \caption{Fig. from Ref. \cite{koch2020adiabatic}. Left: strings' pattern in the complex plane and its interpretation as bound states. Right: bound state production at $c=0^-$ for different $c=0^+$ root densities. Since the quations are diagonal in the rapidity space, we show the bound state production for a given $\rho(\lambda)$. The low density limit shows clear differences from the exact result as $\rho(\lambda)$ is increased}
    \label{fig:LL_bound}
\end{figure}

Even though the model (in the absence of the trap) is integrable for arbitrary values of $c$, it features two completely different phases depending on its sign.
On the contrary of the repulsive regime $c>0$, in the attractive phase $c<0$ the Bose Gas can form bound states of an arbitrary number of particles \cite{doi:10.1063/1.1704156}, hence the hydrodynamic description needs infinitely many root densities $\{\rho_j(\lambda)\}_{j=1}^\infty$, where $\rho_j(\lambda)$ describes the population of the bound state, or string, of $j-$particles (see Fig. \ref{fig:LL_bound}).
More specifically, the solutions of the Bethe-Takahashi equations form a regular pattern in the complex plane, with groups of rapidities sharing the same real part, but regularly shifted in the imaginary direction $\{\lambda-ic(j+1-2a)/2\}_{a=1}^j$. These groups of rapidities are called strings and describe a bound state of $j$ particles.
Clearly, the point $c=0$ divides two deeply different regions in the phase space: this is pointed out by the thermodynamic instability of the ground state \cite{doi:10.1063/1.1704156,PhysRevA.73.033611} and of finite-temperature thermal states within the attractive phase. Indeed, the homogeneous ground state energy scales over-extensively with the number of particles $E_{GS}\propto -N^3$ making impossible to define an intensive energy density and hence thermodynamics in the usual sense.
This can be easily seen from the energy eigenvalues of the bound states
\be\label{eq_en_attractive_bosegas}
\epsilon_j(\lambda)=j\lambda^2-\frac{c^2}{12}j(j^2-1)\, ,
\ee
which grow up to arbitrarily negative values increasing $j$. Therefore, on thermal states both at finite and zero temperature, the gas prefers to form a single giant bound state.
Out-of-equilibrium, this instability can be resolved, thanks to the fact that integrability prevents the system to thermalize, hence to become unstable. This has been already understood in sudden interaction quenches from the repulsive to the attractive regime, even though analytical results were available only in a handful of cases \cite{PhysRevLett.116.070408,SciPostPhys.1.1.001}.
On the contrary, here we wish to apply the hydrodynamic description and are therefore interested in slow variations of the interactions. For the time being, we focus on homogeneous systems with a time-dependent interaction $c\to c(t)$: spatial inhomogeneities can be also included and will be briefly discussed afterwards.

If one remains within either the repulsive or attractive regime, the system falls within the hydrodynamics we already discussed in Sec. \ref{sec_force_first_order}: the root density, or equivalently the filling, evolves under an effective force due to interaction's changes Eq. \eqref{eq_f1}.
The forces \eqref{eq_f1} are computed from the scattering data, which for the repulsive phase are reported in Eq. \eqref{eq_ScM_LL}, while in the attractive phase are
\be\label{eq_scattering_phase_bose_attractive}
\Theta_{j,k}(\lambda)=(1-\delta_{j,k}\theta_{|j-k|}(\lambda)+\theta_{j+k}(\lambda)+2\sum_{\ell}^{\text{min}(j,k)}\theta_{|j-k|+2\ell}(\lambda)\, ,\hspace{1pc} \theta_j(\lambda)=-2\arctan(2\lambda/(jc))
\ee
with the momentum eigenvalue $p_j(\lambda)=j\lambda$ and the energy eigenvalues already anticipated in Eq. \eqref{eq_en_attractive_bosegas}. The repulsive case has been already discussed in Section \ref{sec_firstorder_lackth}; we notice that the repulsive scattering data are obtained from Eq. \eqref{eq_scattering_phase_bose_attractive} retaining only the first string, and of course choosing a positive value of $c$.

With the methods discussed in Section \ref{sec_force_first_order} one can control the hydrodynamics within the two phases, but in order to have a unified framework it must be understood how to connect them while crossing $c=0$.
Physically, as $c$ approaches zero from below, the bound states becomes shallower and shallower and one can imagine them to be completely indistinguishable from unbound particles: this is the key observation that allows one to cross $c=0$ and connect the $c=0^+$ root density $\rho(\lambda)$ with the $c=0^-$ ones $\rho_j(\lambda)$.
More quantitatively, let us consider the expectation value of a conserved charge $Q$ within the attractive phase
\be
L^{-1}\langle Q\rangle=\sum_j \int \dd\lambda\,  h_j(\lambda)\rho_j(\lambda)\, ,\ee
where with $h_j(\lambda)$ we denote the charge-eigenvalue associated with the $j^\text{th}-$string and is obtained summing over the constituents of the string \cite{takahashi2005thermodynamics}
\be\label{eq_ch_LL}
h_j(\lambda)=\sum_{a=1}^j h_1(\lambda-i c (j+1-2a)/2)\, .
\ee
Let us now assume the charge $Q$ to be continuous in $c=0$, i.e. the charge density is an operator that is continuous in the interaction across zero.
In this case, one expects the charge eigenvalues to be continuous as well $\lim_{c\to 0^-} h_j(\lambda)=j\lim_{c\to 0^-}h_1(\lambda)$.
 
Furthermore, since $h_1(\lambda)$ is computed in the 1-particle sector and a single particle is insensitive to the interactions of the many body Hamiltonian,
we can rightfully assume that in the limit $c\to 0^-$, the charge eigenvalue $h_1(\lambda)$ coincides with the $c\to 0^+$ limit of the charge eigenvalue in the repulsive phase $\lim_{c\to 0^-}h_1(\lambda)=\lim_{c\to 0^+} h(\lambda)$. Combining these facts, in the case of continuous charges one finds
\be\label{eq_continuity_local}
\lim_{c\to 0^+}\langle Q\rangle=\lim_{c\to 0^-}\langle Q\rangle\hspace{1pc}\Longrightarrow\hspace{1pc} \int \dd\lambda \, \Big[\rho(\lambda)-\sum_j j\rho_j(\lambda)\Big]\lim_{c\to 0^+}h(\lambda)=0\, .
\ee
Let us now discuss the continuity of the charges. Strictly local charges, as energy and momentum, can be expected to be continuous through $c=0$.
However, the attractive phase is expected to have quasi-local charges due to the presence of the bound states \cite{Ilievski_2016} that cannot be expected to be continuous. On the other hand,  the range of these charges is connected with the size of the bound states $\propto 1/|c|$ and therefore they cannot be included in a GGE with a finite correlation length. Hence, quasilocal charges are discarded and the completeness of the local ones constrains the state through \eqref{eq_continuity_local}
\be\label{eq_root_continuity}
\rho(\lambda)=\sum_j j \rho_j(\lambda)\, ,
\ee
which we stress to be diagonal in the rapidity space.
We notice \eqref{eq_root_continuity} determines the solution if one is moving from the attractive to the repulsive phase: in this case, $\{\rho_j(\lambda)\}_{j=1}^\infty$ are known and $\rho(\lambda)$ is the variable to be determined.
On the opposite scenario when one enters in the attractive phase from the repulsive one, Eq. \eqref{eq_root_continuity} is not sufficient: we will be back to this problem after having discussed the physical meaning of the above equation.
 Indeed, the fact that Eq. \eqref{eq_root_continuity} is diagonal in the rapidity space can be understood in terms of very simple semiclassical arguments. If one moves from $c=0^-$ to $c=0^+$, a bound state of particles is melted in a group of unbound particles moving with the same velocity and hence with the same rapidities. Indeed, at the non interacting point is easy to see $v^\text{eff}(\lambda)=v^\text{eff}_j(\lambda)=2\lambda$.
Let us now consider the opposite scenario, when the interaction is changed from positive to negative values. In this case, only excitations that remain close for a long time can be bound together by the short ranged interactions, hence they must share the same velocity: excitations with different velocities will eventually drift apart before forming a bound state.

Changing the interaction from positive to negative, Eq. \eqref{eq_root_continuity} does not uniquely determine the $c=0^-$ state, but physical arguments discussed before tell us bound states are indistinguishable, hence there is not a preferred way to form them.

Therefore, in the usual spirit of GGEs, $\rho_j(\lambda)$ are determined by entropy maximization constrained to Eq.\eqref{eq_root_continuity}.

In practice, $\rho_j(\lambda)$ is found maximizing the free energy $F$
\be
F=\int \dd\lambda\Big\{ -\sum_j\rho_j^t(\lambda)\big[n_j(\lambda)\log n_j(\lambda)+(1-n_j(\lambda))\log(1-n_j(\lambda))\big]-\omega(\lambda)\sum_j j\rho_j(\lambda) \Big\}\, ,
\ee
where $\omega(\lambda)$ is a rapidity-dependent Lagrange multiplier, which is fixed by imposing Eq. \eqref{eq_root_continuity}.
The free energy maximization is performed through standard TBA manipulations \cite{takahashi2005thermodynamics}, which are further simplified due to the fact that in the $c\to 0$ limit the dressing equations become diagonal in the rapidity space 
\be
\rho_j^t(\lambda)=\frac{j}{2\pi}-\sum_{k}(2\min(j,k)-\delta_{j,k})\rho_k(\lambda)\, .
\ee
The free energy maximization results in the following equations \cite{koch2020adiabatic}
\be\label{eq_max_entr}
\varepsilon_j(\lambda)=j\omega(\lambda)+\sum_k (2\min(j,k)-\delta_{j,k})\log(1+e^{-\varepsilon_k(\lambda)})\, ,
\ee
where, as customary, we parametrize the filling in terms of the effective energy $n_j(\lambda)=(1+e^{\varepsilon_j(\lambda)})^{-1}$. 
Interestingly, we notice that despite the focus on the non-interacting point $c=0$, the equations \eqref{eq_max_entr} captures the effect of strong correlations due to the presence of the sum on the r.h.s..
Indeed, Eq. \eqref{eq_max_entr} can be checked against microscopic calculations in the low density limit $\rho(\lambda)\to 0$ by solving an infinitesimal quench in the attractive phase: this results in $\varepsilon_j(\lambda)=j\omega(\lambda)$, i.e. the leading low-density limit of Eq. \eqref{eq_max_entr}.

A word of caution should be spent on the role of a having a finite-temperature in the initial state, which smoothes out the initial Fermi sea: some subtleties arise if one starts from the exact ground state at $c>0$. 
In this case, by solving the GHD equation one follows the instantaneous ground state \cite{DOYON2018570}, forming the BEC condensate $\rho(\lambda)\propto \delta(\lambda)$ at $c=0^+$: in this case, Eq. \eqref{eq_max_entr} predicts that bound states of arbitrary length $j$ are populated and thus the instability of the attractive phase.  Furthermore, the very founding idea of the indistinguishability of bound states from unbound particles does not apply, since the BEC has infinite correlation length.
Having a finite temperature in the initial state regularizes the protocol: in general, initial states with the same density but lower temperature excite lager boundstates.

As we previously mentioned, the approach can be also generalized to the case of spatially inhomogeneous smooth interactions $c\to c(x)$. In this case, we could imagine of smoothly connecting a repulsive half with an attractive one: particles traveling from $c>0$ to $c<0$ will bind together, while bound states that come from the attractive region and enter into the repulsive one, will split in unbound particles.
Besides the per se theoretical interest of this scenario, and its importance for future developments as outlined in Section \ref{eq_BS_perspectives}, new advances in experimental techniques can engineer such inhomogeneous interactions in the 1d Bose gas \cite{PhysRevLett.115.155301}.

On the practical level, the spatially inhomogeneous transition is governed by the same equations that determine the time-dependent transition Eqs. \eqref{eq_root_continuity} and \eqref{eq_max_entr}, albeit they are derived by different means.
Indeed, the continuity of the charges \eqref{eq_continuity_local} is replaced by the continuity of the currents at the interface between the two phases. Furthermore, in place of the entropy maximization one considers the entropy growth: we leave the details to the original reference  \cite{koch2020adiabatic}.
A word of caution should be spent Eqs. \eqref{eq_root_continuity} and \eqref{eq_max_entr} applied to the spatially inhomogeneous case, since the role of the root densities as inputs or outputs of the equations depend on the sign of their velocity, which it determines if they are flowing towards the $c=0$ zone (input) or leaving it (output).

Before of closing this section, we would like to add some extra comments about the entropy. Indeed, within the two phases the entropy is conserved by the eulerian hydrodynamic, but crossing $c=0$ there is room for changes. In principle, since bound states are produced and there are many different ways to arrange their microscopic populations, one could expect that the entropy is increased, but in Ref. \cite{koch2020adiabatic} it has been numerically observed that the equations \eqref{eq_max_entr} conserve the entropy crossing $c=0$. We notice that, since Eq. \eqref{eq_max_entr} has been derived maximizing the entropy, it means that the entropy of any other admissible set of root densities $\{\rho_j\}_{j=1}^N$ would be smaller than entropy of the $c=0^+$ state.
However, entropy conservation is a feature of this model: in the next section we will see how the bound states recombination in the XXZ causes a non trivial evolution of the Yang-Yang entropy.

\subsection{Magnetic flux changes in the XXZ spin chain}
\label{sec_XXZ_flux}

The interacting Bose gas is only one among the integrable models which feature recombination of bound states at a hydrodynamic level. Another example is the XXZ spin chain under the action of a time-dependent magnetic flux \cite{PhysRevLett.122.240606}: here, the recombination of bound states takes place in a much more subtle way.
Let us consider the following Hamiltonian
\be\label{eq_xxz_fluxH}
\hat{H}(\Phi)=\sum_{j=1}^N\frac{1}{2}\left(e^{i\Phi}\hat{s}^+_j \hat{s}^{-}_{j+1}+e^{-i\Phi}\hat{s}^+_{j+1}\hat{s}_j^-\right)+\Delta \hat{s}^z_j \hat{s}^z_{j+1}\, ,
\ee
where $s^{x,y,z}$ are canonical $1/2$ spin operators and $s^\pm=(s^x\pm is^y)/2$ and periodic boundary conditions are assumed. For the sake of simplicity, following Ref. \cite{PhysRevLett.122.240606} we consider a homogeneous scenario.
The variable $\Phi$ describes the flux of a magnetic field piercing the ring of the spin chain. 
In the absence of magnetic flux $\Phi=0$, one gets the standard XXZ spin chain, which is a widely studied integrable model.
On the other hand, the value of the magnetic flux in itself should not change the physics of the system, which is sensible only to the variations of $\Phi$. On the XXZ spin chain, this assertion is reflected on the fact that the Hamiltonian \eqref{eq_xxz_fluxH} is equivalent to the Hamiltonian $H(\Phi=0)$ after a gauge transformation
\be\label{eq_gauge_transformation}
\hat{H}(\Phi)=W^\dagger_\Phi\hat{H}(0)W_\Phi\, ,\hspace{2pc} W_\Phi=\exp\left[-i\Phi\sum_j j \hat{s}_j^z\right]\, .
\ee
The action of $W_\Phi$ on the spin operators $\hat{s}_j^\pm$ simply adds a global phase $W_\Phi^\dagger \hat{s}^+_j W_\Phi=e^{-i j\Phi} \hat{s}_j^+$, while leaves $\hat{s}^z_j$ unscathed. 

We are interested in studying magnetic flux changes, while the interaction $\Delta$ is kept constant: a time-dependent magnetic flux results in an effective electric field along the spin chain. Hence, the excitations are accelerated.

Since the model is instantaneously integrable for arbitrary values of the magnetic flux $\Phi$, in the case of adiabatic flux changes the problem can be studied with GHD. 
Some preliminary comments are due: first, we notice that due to the gauge symmetry the excitation's content of the system, i.e. its TBA description, and the scattering matrix do not change with $\Phi$. 
However, the Hamiltonian and the (quasi-)local conserved quantities change in a non-trivial way under \eqref{eq_gauge_transformation}. As a consequence, even though the description of the model in terms of root densities does not change with $\Phi$, the underlying integrable model does.
Without lack of generality, we can focus on positive values of the interaction $\Delta$: the phase space of the XXZ spin chain greatly differs whenever $0<\Delta<1$ or $\Delta>1$ \cite{takahashi2005thermodynamics}.
In the $\Delta>1$ case, the system features infinitely many strings $\{\rho_j(\lambda)\}_{j=1}^N$ and the rapidity is defined on a compact support $\lambda\in [-\pi/2,\pi/2]$.
As we will see, the interval $[-\pi/2,\pi/2]$ can be interpreted as a Brillouin zone: this is a crucial point for the forthcoming analysis.
In the case $0<\Delta<1$, the TBA has a fractal structure. Within this phase, the interactions are conveniently parametrized as $\Delta=\cos(\pi \gamma)$: for rational values of $\gamma$, the number of strings is finite and depends on the continued fraction representation of $\gamma$, and increases as more terms in the continued fraction representation are needed. Hence, the number of strings is nowhere continuous in $\gamma$.
For the sake of concreteness, we focus on the simplest case of $\gamma=1/\ell$, which is described by exactly $\ell$ strings, but the forthcoming hydrodynamic approach is of more general validity.
In the $0<\Delta<1$ case, the rapidity parametrization covers the entire real axis $\lambda\in(-\infty,\infty)$ and it cannot be interpreted as a Brillouin zone.

The discussion concerning the Brillouin zone must be further clarified. Indeed, the momentum in the discrete model Eq. \eqref{eq_gauge_transformation} always lives in a Brillouin zone $[-\pi,\pi]$. Regardless, the momentum of a given string $p_j(\lambda)$ can cover or not $[-\pi,\pi]$ as $\lambda$ spans its definition domain.
Let us be more concrete and start analyzing the case $\Delta>1$. In this case, one conveniently parametrizes the interaction as $\Delta=\cosh\theta$. The momentum eigenvalue associated with the $j^\text{th}$ string is 
\be\label{eq_p_xxz_deltaM}
p_j(\lambda)=\sum_{a=0}^{j-1}p\left(\lambda+i\theta \frac{j-1-2a}{2}\right)\,,\hspace{2pc}p(\lambda)=-i\log\left(\frac{\sin(\lambda-i\theta/2)}{\sin(\lambda+i\theta/2)}\right)\, .
\ee

One can easily check that when $\lambda$ is changed within interval $[-\pi/2,\pi/2]$, the momentum $p_j(\lambda)$ covers the entire Brillouin zone $[-\pi,\pi]$. Furthermore, choosing the branch cut of the logarithm in such a way to be continuous in $\lambda=0$, one finds $p_j(\lambda=\pi/2)=p_j(\lambda=-\pi/2)+2\pi$. We can then interpret $\lambda\in[-\pi/2,\pi/2]$ as a Brillouin zone in the rapidity space
\footnote{More precisely, one should check the periodicity in $\lambda$ of the whole microscopic wavefunction.}.
Let us now consider the other phase, with $\Delta=\cos(\pi\gamma)$.
In this case the momentum is
\be\label{eq_p_xxz_deltam}
p_j(\lambda)=\sum_{a=1}^{m_j}p\left(\lambda+i\gamma \frac{m_j+1-2a}{2}+i\frac{\pi(1-\xi_j)}{4}\right)\,,\hspace{2pc}p(\lambda)=-i\log\left(\frac{\sinh(\lambda+i\gamma/2)}{\sinh(\lambda-i\gamma/2)}\right)\, .
\ee
The number of strings, the magnetization eigenvalue $m_j$ and $\xi_j$ depend on the value of $\gamma$. In the case $\gamma=1/\ell$ one has $\ell$ strings and
\be\label{eq_mag_eigenv}
m_j=j\,,\hspace{1pc} \xi_j=1\hspace{1pc}\text{for } j< \ell\hspace{1pc}\text{and}\hspace{1pc}
m_\ell=1\,,\hspace{1pc} \xi_\ell=-1
\ee
We can readily see that
\be\label{eq_limrapinf}
\lim_{\lambda\to \pm \infty}p\left(\lambda+i\gamma \frac{m_j+1-2a}{2}+i\frac{\pi(1-\xi_j)}{4}\right)=\pm \pi\gamma
\ee

The total momentum $p_j$ is not periodic modulus $2\pi$ over $\lambda\in (-\infty,\infty)$ since $\lim_{\lambda\to\pm \infty}p_j(\lambda)=\pm \pi \gamma m_j$ and $2\pi \gamma m_j\ne 0\mod 2\pi$.

Interestingly, the momentum of each of the components of a given string reaches the same value at $\lambda\pm \infty$: this can be argued for all the other local conserved charges as well \cite{PhysRevLett.122.240606} and suggests that the strings are ``connected" at $\lambda=\pm \infty$.
In other words, an excitation sat at $\lambda=\pm \infty$ can be equivalently thought to belong to any string, similarly to what happened in the interacting Bose gas of Section \ref{eq_sec_attractive_bose_gas} where, at $c=0^-$, the bound states could not be distinguished from unbound particles.

Before of focusing on the GHD equations, it is useful to qualitatively point out the difference between the phase $\Delta>1$ and $0<\Delta<1$.
The change in the magnetic flux will exert a force on the excitations, which are then accelerated changing their rapidities. Imagine we accelerate an excitation to the edges of the rapidity domain. In the $\Delta>1$, where one has a Brillouin zone, one can further accelerate the excitation bringing it back to the other edge of the rapidity domain, in a sort of Bloch oscillation.
On the other hand, in the $0<\Delta<1$ this is not allowed, but our previous experience with the Bose gas and Eq. \eqref{eq_limrapinf} gives a hint: the excitation is moved to another string where it keeps accelerating.
Hence, we have bound state recombination at the edges of the rapidities domain, but only in the $0<\Delta<1$ case.

The remaining part of this section is devoted to make this analysis quantitative through the hydrodynamic analysis.
Focusing on the homogeneous case, one obtains the simple hydrodynamic equations \cite{PhysRevLett.122.240606}
\be\label{eg_ghd_flux}
\partial_t n_j(\lambda)+\partial_t\Phi\frac{m_j^\text{dr}}{(\partial_\lambda p_j)^\text{dr}}\partial_\lambda n_j(\lambda)=0
\ee
 
It is worth to be said that, while for $\Delta>1$ one has $\partial_\lambda p_j>0$, in for $0<\Delta<1$ the sign of the derivative of the momentum depends on the string. Hence, the relation between the dressed momentum and the total root density, which must be positive, is modified  \cite{takahashi2005thermodynamics}
\be\label{eq_roott_flux}
(\partial_\lambda p_j)^\text{dr}=2\pi \sigma_j \rho_j^t(\lambda)\, .
\ee

\begin{figure}
    \centering
    \includegraphics[width=0.8\textwidth]{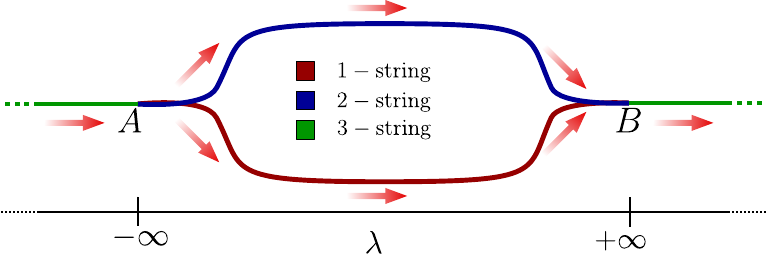}
    \caption{Pictorial representation of the excitation's motion in the string space for $\Delta=0.5$. Assuming $\partial_t\Phi>0$, the first and second string are right movers in the rapidity space, while the $3^\text{rd}$ string is a left mover. Let us for example consider an excitation originally places in the first string: this will be accelerate until $\lambda=+\infty$ (point B) where it becomes an excitation of the $3-$string. Then, it will decrease its rapidity to $-\infty$ (point A). Here, it can continue as a $1-$string or $2-$string: the probability of picking one choice rather than the other is determined on the basis of entropic considerations.}
    \label{fig:flux2}
\end{figure}

In the case at hand, i.e. $\gamma=1/\ell$, $\sigma_{j<\ell}=1$ and $\sigma_\ell=-1$.
Let us now consider a linear increasing of the flux $\partial_t\Phi>0$:

In the $\Delta>1$ case, $\partial_\lambda p_j>0$ for any string, hence the fillings are coherently moving in the same direction according to Eq. \eqref{eg_ghd_flux}. 
In the other phase $0<\Delta<1$, the sign of $\partial_\lambda p_j$ depends on the string: strings with opposite signs in $\sigma_j$ move in opposite directions in the rapidity space.
For example, in the case $\gamma=1/\ell$ and for $\partial_t\Phi>0$, the first $(\ell-1)^\text{th}$ strings move to the right, while the $\ell^\text{th}$ string moves to the left, see Fig. \ref{fig:flux2}.
So far, we kept the discussion at a qualitative level: we will now find the equations that determine the boundary conditions at $\lambda=\pm \infty$.
In order to be quantitative, we take a step back in the derivation of the GHD equations presented in Section \ref{sec_force_first_order} and have a better look at Eq. \eqref{eq_intchc}, here reported in the case of multiple strings and $f_j(\lambda)=m_j$ \eqref{eq_intchc}
\be\label{eq_intchc2}
\sum_j\int_{-\infty}^{+\infty}\dd\lambda\, h_{i,j}^{(\Phi)}(\lambda) \partial_t\rho_j(\lambda)=
\partial_t\Phi\sum_j\int_{-\infty}^{+\infty} \dd\lambda\, \frac{m_j^\text{dr}(\lambda)}{(\partial_\lambda p_j)^\text{dr}}\partial_\lambda h_{i,j}^{(\Phi)}(\lambda) \rho_j(\lambda)\, .
\ee
In Section \ref{sec_force_first_order}, we integrated by parts the r.h.s. assuming that boundary terms do not matter, but in the case at hand they give an important contribution. 
By imposing the continuity of the charges $\lim_{\lambda\to +\infty} h^{(\Phi)}_{i,j}(\lambda)=m_j\lim_{\lambda\to+\infty} h^{(\Phi)}_{i,1}(\lambda)$ similarly to what we discussed for the Bose gas,  from the boundary terms one finds the analogue of Eq. \eqref{eq_root_continuity} 
\be\label{eq_xxz_continuity}
\sum_j \sigma_j m_j^\text{dr}(\lambda) m_j n_j(\lambda)\Bigg|_{\lambda=+\infty}=0\, .
\ee
Comparing the above condition with the GHD equations, one is clearly enforcing the current conservation in the rapidity space. It is instructive to use the explicit values of $\sigma_j$ in the above, where one finds $\sum_{j=1}^{\ell-1}m_j^\text{dr}(\lambda) m_j n_j(\lambda)|_{\lambda=+\infty}= m^\text{dr}_\ell(\lambda) m_\ell n_\ell(\lambda)|_{\lambda=+\infty}$, which clearly states that the first $(\ell-1)^\text{th}$ strings flow in the last one and the other way around, depending on the sign of the flux change.
Analogously to the interacting Bose gas of Section \ref{eq_sec_attractive_bose_gas}, the continuity equation is not enough to uniquely determine the recombination of the bound states. In this case, looking at the strings as indistinguishable, we reside to entropic arguments.
Let us consider the Yang-Yang entropy and compute its rate $\partial_t S$, which is zero except for boundary terms $\partial_t S=\partial_t S^++\partial_t S^-$
\be
\partial_t S^\pm=\mp\partial_t\Phi \lim_{\lambda\to\pm\infty}\left[\sum_j \sigma_j m_j^\text{dr}(-n_j\log n_j-(1-n_j)\log(1-n_j))\right]\, .
\ee
Since the strings are indistinguishable at $\lambda=\pm \infty$, they are expected to rearrange maximizing the entropy growth and the following equations are obtained
\be\label{eq_maxentropy_flux}
\log\left(\frac{n_j}{1-n_j}\right)=\omega m_j^\text{dr}-\sum_{j,j'}\frac{1}{2\pi}\tilde{\varphi}_{j,j'}\sigma_{j'}(-n_{j'}\log n_{j'}-(1-n_{j'})\log(1-n_{j'}))\,,
\ee
where, for infinite rapidities, the dressing can be replaced with the simple matrix equation (all the quantities are computed at $\lambda=\pm\infty$)
\begin{figure}
    \centering
    \includegraphics[width=0.5\textwidth]{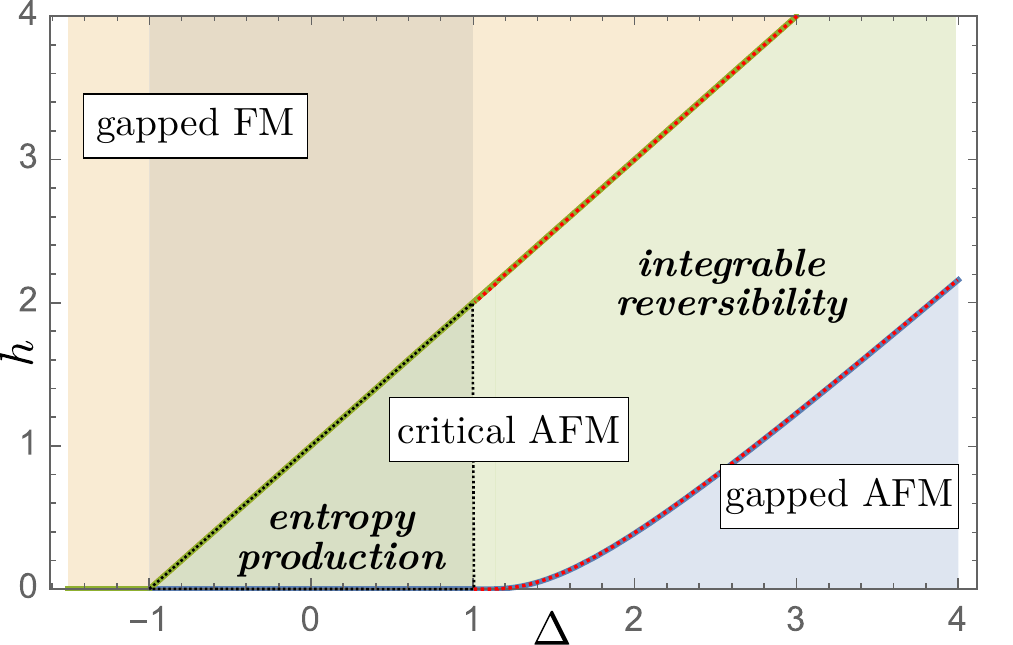}
   \caption{From \cite{PhysRevLett.122.240606}. Phase diagram of the XXZ spin chain under flux changes, assuming the initial state is the absolute ground state of the chain in an external magnetic field $
\hat{H}=\sum_{j=1}^N\frac{1}{2}\left(\hat{s}^+_j \hat{s}^{-}_{j+1}+\hat{s}^+_{j+1}\hat{s}_j^-\right)+\Delta \hat{s}^z_j \hat{s}^z_{j+1}+B\hat{s}_j^z$.}
    \label{fig:flux_phasediagram}
\end{figure}

\be
m_j^\text{dr}=m_j-\sum_{j'}\frac{1}{2\pi}\tilde{\varphi}_{j,j'}\sigma_{j'}m^\text{dr}_{j'}\, ,\hspace{2pc}
\tilde{\varphi}_{j,j'}=\int_{-\infty}^{+\infty}\dd\lambda\,  \varphi_{j,j'}(\lambda)\, .
\ee
where $\varphi_{j,j'}$ is, as usual, the rapidity derivative of the scattering phase, see Ref. \cite{PhysRevLett.122.240606} for the explicit expression.
The Lagrange multiplier $\omega$ is needed to enforce Eq. \eqref{eq_xxz_continuity}.
In Eq. \eqref{eq_maxentropy_flux}, the index $j$ does not run over all the strings, but only on those associated with the fillings $n_j$ to be determined. Hence, if $\partial_t\Phi>0$, strings such that $\sigma_j=1$ flow at the boundary $\lambda=+\infty$, while strings with $\sigma_j=-1$ emerge from the boundary $\lambda=+\infty$ and must be determined. Looking at $\lambda=-\infty$ the role of the signs is of course reversed.

In contrast with the interacting Bose gas discussed in Section \ref{eq_sec_attractive_bose_gas}, in this case the entropy rate maximization causes a non-trivial entropy rate. As a consequence, if one starts from the ground state at fixed magnetization and keep on increasing the flux, higher strings are eventually populated and the system approaches an infinite-temperature state at later times \cite{PhysRevLett.122.240606}, which of course it is not the case for $\Delta>1$, in view of entropy conservation.

The difference between the two phases $\Delta>1$ and $0<\Delta<1$ can also be traced in the (lack of) reversibility of flux changes. Indeed, in the homogeneous GHD equations \eqref{eg_ghd_flux} are \emph{i)} explicitly time-reversal invariant and \emph{ii)} can be parametrized in terms of the flux regardless its time dependence. Therefore, if one first increases the flux up to a certain value, then changes it back to the starting point, would go back to the original state. This reasoning can be applied to the case $\Delta>1$, since both the GHD equations and the boundary conditions at $\lambda=\pm \pi/2$ are invariant under time reversal. However, this is not the case for $0<\Delta<1$, due to the boundary conditions at $\lambda=\pm \infty$ which break the time reversibility of the protocol.
The phase diagram of the model is summarized in Fig. \ref{fig:flux_phasediagram}.

\subsection{Perspectives}
\label{eq_BS_perspectives}

Even though smooth changes of the interactions can deeply affect the integrability structure of certain models, one can still rely on a hydrodynamic approach to capture the evolution of the systems. Despite the fact that the problem has only been addressed on a case by case basis, the two examples studied so far provide some hints on the general scenario. As the interaction changes, the excitations of one phase are smoothly changed into excitations of the other, often allowing for multiple choices which are nevertheless compatible with the set of conserved quantities.
After imposing charge conservation, since there are no any other means to distinguish among the excitations, these get redistributed according to entropic arguments, in an attempt to maximally redistribute in the available phase space.
Beside being of utmost theoretical interest, excitation's recombination is expected to be relevant for several experimental setups.
For example, interaction changes in the Bose gas are within the reach of the state-of-the-art cold atoms experiments. In Ref. \cite{Haller1224}, sudden interaction quenches from the repulsive to the attractive phase have been explored, but more recently the interest turned towards adiabatic protocols moving across the two phases \cite{Kao296}.

The XXZ spin can be engineered in the lab as well \cite{Jepsen2020} and, besides magnetic flux changes, bound state-recombination should be expected every time a force is able to accelerate the particles to the edges of the rapidity domain. This can also be induced by mean of weak gradients of the magnetic field along the $z-$direction, which acts as a inhomogeneous chemical potential for the underlying excitations.

Several interesting questions and models are yet to be addressed. For example, interaction changes in the XXZ spin chain could hide exciting phenomena: in the planar regime $0<\Delta<1$, the excitations' content is nowhere continuous in $\Delta$ and this severely challenges the hydrodynamic approach. Does this lack of continuity prevent one from using GHD? The answer to this question passes through a careful study of how the excitations are transformed varying $\Delta$ and whether it is possible to re-establish  a notion of continuity. This discontinuous excitation spectrum also has interesting consequences for spin transport as it leads to a fractal Drude weight~\cite{PhysRevLett.106.217206,PhysRevLett.111.057203} and anomalous a.c.~conductivity~\cite{2019arXiv190905263A}, as discussed in the reviews on linear response and  anomalous transport~\cite{2021arXiv210301976B} in the same special issue. 
Another model of outstanding experimental relevance is the sine-Gordon (SG) model
\be\label{eq_SG_H}
H=\int \dd x\, \frac{v\pi}{K} \Pi^2+\frac{vK}{4\pi}(\partial_x\phi)^2-\Delta \cos\phi\, ,
\ee
with $\phi$ and $\Pi$ being conjugated fields $[\phi(x),\Pi(y)]=i\delta(x-y)$.
The sine-Gordon emerges as the low-energy theory of two weakly-coupled 1d Bose gases, where a small tunneling is allowed between the two tubes
\cite{PhysRevB.75.174511}. In this setup, the parameters $v$ and $K$ in the Hamiltonian \eqref{eq_SG_H} are the sound velocity and $K-$Luttinger parameter of the two atom clouds in the two tubes, while $\Delta$ is proportional to the tunneling.
This setup has been thoroughly experimentally studied \cite{Gring1318,Schweigler2021} and theoretically modeled through different approaches \cite{PhysRevLett.121.110402,van_Nieuwkerk_2019,10.21468/SciPostPhys.9.2.025,vannieuwkerk2020josephson}, but a GHD treatment of the realistic experimental apparatus has yet to be attempted. 
The presence of the longitudinal trapping for the atomic clouds induces a weak inhomogeneity in the sound velocity $v$ and Luttinger parameter $K$, the latter reaching the non-interacting limit $K=1$ at the edges of the trap and the deep semiclassical limit in the center $K\sim 70$ \cite{Gring1318}.
The value of $K$ is extremely important in identifying the excitations' content of the theory. Indeed, besides the solitonic modes which are present for any $K$, the SG model features bound states, called breathers, whose number depends on $K$ as the integer part of $4K-1$. The changes in $K$ along the longitudinal trap are expected to cause breathers to split in solitons and vice versa: this problem has yet to be addressed.

\section{Conclusions and outlook}
\label{sec_conclusions}

In this review, we have described recent progress in the understanding of the hydrodynamics of systems close to integrability, building on the framework of GHD. The question of integrability breaking is both of fundamental and practical interest, as realistic systems are never perfectly integrable. 
Qualitatively, the effects of typical integrability breaking perturbations are rather clear: (1) Inhomogeneous potentials lead to generalized force terms than accelerate quasiparticles, and (2) at second and higher order in perturbation theory, integrability-breaking perturbations can cause quasiparticles to scatter, leading to thermalization and chaotic dynamics at long times. For example, a realistic 1d Bose gas with integrability breaking perturbations conserving particle number, momentum and energy is expected to be described by conventional Navier-Stokes hydrodynamic equations at long times, while shorter time scales (still in the hydrodynamic regimes) are captured by GHD. Similarly, spin chains near integrability should in general exhibit diffusive hydrodynamic transport of their conserved quantities (energy or spin) at long times.
However, capturing this crossover from generalized to conventional hydrodynamics quantitatively, and computing transport coefficients perturbatively remains a formidable task, with dramatic progress over the past couple of years enabled by the theory of GHD. 

As we have described in section~\ref{sec_inh_couplings}, the leading order effects of inhomogeneous potentials are now relatively well understood, though questions remain regarding higher-order corrections and integrable systems with discontinuous quasiparticle spectrum. Meanwhile, the framework of GHD is a natural starting point to develop a generalized Boltzmann equation formalism for strongly interacting perturbed integrable systems, as explained in section~\ref{sectionBoltzmann}. There, the transport coefficients (diffusion constants) of the perturbed model can be expressed in terms of GHD data, and of ``collision integrals'' that depend on matrix elements of the perturbations. A lot of challenges remain in this direction, from more explicit expressions of those matrix elements through the form factor program, to developing more accurate and general approximations of collision integrals. Numerical benchmarks and checks, as well as applications to realistic experimental setups also represent natural directions for future research. We expect that developing new numerical techniques that can capture the dynamics of thermalizing systems efficiently will also be crucial in that respect (see Refs.~\cite{2017arXiv170208894L,PhysRevB.97.035127,2020arXiv200405177R} for recent progress in this direction).
Finally, we have also discussed hidden non-adiabadic effect and bound states recombination in integrable systems with discontinuous spectrum, whose physics are closely related to integrability breaking. Multiple open questions in this direction were summarized in section~\ref{eq_BS_perspectives}. 

An especially exciting research direction is to identify perturbations that only ``weakly'' break integrability, in the sense that they would lead to vanishing collision integrals at a given order in perturbation theory. More generally, some perturbations could allow for arbitrarily long-lived quasiparticles and anomalous transport properties in non-integrable systems, at least up to very long time scales (see {\it e.g.}~\cite{2020arXiv200411030D,2021arXiv210202219D,PhysRevLett.125.030601}  for recent results along those lines). 
In general, we expect that applications of the GHD framework to many-body systems close to integrability has only just started, and we anticipate numerous developments in the coming years. 

\paragraph{Acknowledgments:} The authors thank  Utkarsh  Agrawal, Vincenzo Alba, Sounak  Biswas, Vir Bulchandani,  Jean-S\'ebastien Caux, Jacopo De Nardis, Benjamin Doyon, Michele Fava, Aaron Friedman, Paolo Glorioso, Sarang Gopalakrishnan, David Huse, Enej Ilievski, Vedika Khemani, Rebekka Koch, Javier  Lopez-Piqueres, Joel Moore, Vadim Oganesyan, Sid Parameswaran, Tomaz Prosen, Marcos  Rigol, Subir Sachdev,  Brayden  Ware and Marko Znidaric for collaborations and/or numerous discussions on topics related to integrability breaking and GHD. R.V. acknowledges partial support from the Air Force Office of Scientific Research under Grant No.~FA9550-21-1-0123, and the Alfred P. Sloan Foundation through a Sloan Research Fellowship.
AB acknowledges support from the Deutsche Forschungsgemeinschaft (DFG, German Research Foundation) under Germanys Excellence Strategy–EXC–2111–390814868.

\appendix

\section{Numerical methods for the GHD equations}
\label{sec_num}

The appeal of GHD resides in the possibility of writing exact hydrodynamic equations able to describe the system's evolution on large space and time scales. However, apart from exceptional cases, the hydrodynamic equations need to be solved numerically.
In this short appendix we present some stable numerical methods that can be used to solve those equations, and how they can be extended to deal with non-integrable settings.
At the moment, a public library is available \cite{10.21468/SciPostPhys.8.3.041} which is highly optimized and customizable, nevertheless we think it is important to give the reader an overview of the possible methods.

We start by presenting the method of the characteristics, implemented in Ref. \cite{10.21468/SciPostPhys.8.3.041}, which is suitable for the Eulerian hydrodynamic equation. This method cannot be applied in the presence of diffusion, that will be discussed afterwards.
Let us consider the Eulerian hydrodynamic equation expressed in terms of the filling function \eqref{eq_ghd_1ord_filling}. In what follows, we fully specify the dependence in time, space and rapidity $n\to n(t,x,\lambda)$. Eq. \eqref{eq_ghd_1ord_filling} is nothing else than an infinitesimal translation of the filling in the phase space.
As such, it has the following implicit solution
\be
n(t,x,\lambda)=n(t_0,x( t,t_0),\lambda(t,t_0)),
\ee
where $x( t,t_0)$ and $\lambda( t,t_0)$ are the backwards evolution of the target coordinates under the generalized equations of motion
\begin{eqnarray}\label{eq_char_x}
&x( t,t_0)=x-\int_{t_0}^{t}\dd t'\, v^\text{eff}(t',x(t',t_0),\lambda(t',t_0)), \, \\\label{eq_char_rap}
&\lambda(t,t_0)=\lambda-\int_{t_0}^{t}\dd t'\, F^\text{eff}(t',x(t',t_0),\lambda(t',t_0)).
\end{eqnarray}

Seeing the time evolution as a shift in the phase space provides good stability: for example, it keeps the filling bounded between $0$ and $1$.
Eqs. \eqref{eq_char_x} and \eqref{eq_char_rap} are only an implicit solution, since the effective velocity and force have a complicate dependence on the evolving state $t'\in(t_0,t)$ due to the dressing.
However, their approximate solution provides useful approximation schemes: let us consider an infinitesimaly evolution $t=t_0+\Delta t$, then at first order in $\Delta t$ one can use a forward Euler method 
\begin{align}
x( t_0+\Delta,t_0)&=x-\Delta t\,  v^\text{eff}(t_0,x,\lambda)+\mathcal{O}(\Delta t^2)\,, \\
\lambda(t_0+\Delta t,t_0)&=\lambda-\Delta t\, F^\text{eff}(t_0,x,\lambda)+\mathcal{O}(\Delta t^2)\, ,
\end{align}
or a backward Euler method
\begin{align}
x( t_0+\Delta t,t_0)&=x-\Delta t\,  v^\text{eff}(t_0+\Delta t,x( t_0+\Delta t,t_0),\lambda(t_0+\Delta t,t_0))+\mathcal{O}(\Delta t^2)\, , \\ 
\lambda(t_0+\Delta t,t_0)&=\lambda-\Delta t\, F^\text{eff}(t_0+\Delta t,x( t_0+\Delta t,t_0),\lambda(t_0+\Delta t,t_0))+\mathcal{O}(\Delta t^2)\, .
\end{align}

This backward Euler method has been introduced for the first time in Ref. \cite{PhysRevLett.119.220604} and it has been observed to have enhanced stability when compared with the forward strategy. As a downside, the backward method is implicit and it has to be iteratively solved, which can be numerically demanding since at each step the integral equations defining the dressing must be solved.
The method of characteristics can be easily improved to second order in the time step \cite{PhysRevLett.123.130602}. Indeed, the first order method comes from approximating the integrals in Eqs. \eqref{eq_char_x} and \eqref{eq_char_rap} with the endpoints of the integration region, while using the middle point increases the order of the approximation
\begin{align}
x( t_0+\Delta t,t_0)&=x-\Delta t\,  v^\text{eff}(t_0+\Delta t/2,x( t_0+\Delta t/2,t_0),\lambda( t_0+\Delta t/2,t_0))+\mathcal{O}(\Delta t^3)\,, \notag \\ 
\lambda(t_0+\Delta t,t_0)&=\lambda-\Delta t\, F^\text{eff}(t_0+\Delta t/2,x( t_0+\Delta t/2,t_0),\lambda( t_0+\Delta t/2,t_0))+\mathcal{O}(\Delta t^3)\, ,
\end{align}
This is not an explicit method yet, since the coordinates at the middle points are not known. However, we can use the first order forward Euler method to estimate up to the first order in $\Delta t$ the coordinates inside the effective force and velocity. This does not change the overall order of the method.
We notice that, while with the first order method one only needs to know the state (i.e. the filling fraction) at a given time $t$ in order to reach the time $t+\Delta t$, the second order algorithm requires the both the states at $t$ and $t+\Delta t/2$, doubling the memory cost.

In both the first and second order methods, one should distinguish between what is referred to as the ``finite-entropy case", where the filling function is a smoothly-varying function, and the ``zero-entropy case". 
In this second case, similarly to what happens in the ground state, the filling function takes only the values $0$ or $1$, abruptly passing from one to the other \cite{PhysRevLett.124.140603}.
In the finite entropy case, one discretizes the phase space on a grid and the translations of the filling are implemented by interpolations (see eg. Ref. \cite{PhysRevLett.119.220604,PhysRevLett.123.130602,10.21468/SciPostPhys.5.5.054}). On the other hand, in the zero entropy situation the filling is a sharp, time evolving step-function and interpolations do not work. In this case, one identifies a ``Fermi contour" in the phase space dividing the regions where the filling is $1$ from those where it is zero, then follows the evolution of this contour \cite{PhysRevLett.123.130602}.
It should be mentioned that the method of the characteristics can be easily generalized to include explicit collisional terms discussed in Section \ref{sectionBoltzmann}.
Let us consider an equation in the following form
\be
\partial_t n+v^\text{eff}\partial_x n+F^\text{eff}\partial_\lambda n=\mathcal{I}_\lambda[n]
\ee
with $\mathcal{I}_\lambda[n]$ a certain functional of the filling function at time $t$. 
This equation admits the following implicit solution
\be
n(t,x,\lambda)=n(t_0,x( t,t_0),\lambda(t,t_0))+\int_{t_0}^t\dd t'\,\mathcal{I}_\lambda[n(t')]\, ,
\ee
where in the collisional integral the filling is computed at time $t'$, we leave the other coordinates unspecified for the sake of simplicity. Similarly to the case without collisional term, the above expression can be discretized at first order with a forward or backward Euler scheme, or at second order considering the middle point. In Ref. \cite{2020arXiv200513546L} the first order backward Euler method has been used to explicitly handle collisional terms.

While discussing the numerical solution of the Eulerian equations, the ``flea-gas method" \cite{PhysRevLett.120.045301} is worth mentioning. The idea is to engineer a gas of classical particles with pairwise elastic scattering, but with momentum dependent time delay. The classical time delay is then matched with the scattering phase of the target model and the initial conditions likewise engineered.
The resulting model has the same Eulerian hydrodynamics of the quantum one \cite{PhysRevLett.120.045301} and acts like a stable simulator. However, to this date it has not been understood how to include force terms or diffusive corrections while properly accounting for the dressing.

We finally discuss the numerical solution of the hydrodynamic equations in the presence of diffusion. In Eq. \eqref{eq:chargediffinho} we wrote the diffusive equation in the basis of the charges, but in practice it is more convenient to write them in the rapidity basis, as a hydrodynamic equation for the root density \cite{PhysRevLett.125.240604}
\be\label{eq_diff_root}
\partial_t \rho(t,x,\lambda)+\partial_x(v^\text{eff}\rho(t,x,\lambda))-\partial_\lambda(\partial_x V \rho(t,x,\lambda))-\int d\lambda'\, \partial_x(\mathfrak{D}_{\lambda}^{\,\lambda'}\partial_x \rho(t,x,\lambda'))=0
\ee
Above, $\mathfrak{D}_{\lambda}^{\,\lambda'}$ is the diffusion matrix in the rapidity space, which explicitly reads (we use a matrix notation, where intergration over repeated indexes is implicit)
\be
\mathfrak{D}_{\lambda}^{\, \lambda'}=[R^{-1}]_\lambda^{\, \mu}\tilde{\mathfrak{D}}_{\mu}^{\, \mu'}R_{\mu'}^{\, \lambda'}\,,\hspace{1pc} \tilde{\mathfrak{D}}_\lambda^{\,\lambda'}=\delta(\lambda-\lambda')\int d \mu \left(\frac{(\partial_\gamma p)^\text{dr}}{(\partial_\mu p)^\text{dr}}\right)^2 W_{\mu}^{\, \lambda'}-W_\lambda^{\, \lambda'}.
\ee

The function $W_\lambda^{\, \lambda'}$ is defined as
\be
W_{\lambda}^{\, \lambda'}=\frac{1}{2}( [\varphi^\text{dr}]_{\lambda}^{\lambda'}/(\partial_\lambda p)^\text{dr})^2 \rho(\lambda)(1-n(\lambda))|v^\text{eff}(\lambda)-v^\text{eff}(\lambda')|,
\ee
with
\be
R_{\lambda}^{\, \lambda'}= \left(2\pi\delta(\lambda-\lambda')+n(\lambda) \varphi_\lambda^{\lambda'}\right)/(\partial_\lambda p)^\text{dr},
\ee
and  
\be
[\varphi^\text{dr}]_{\lambda}^{\, \lambda'}=\varphi(\lambda-\lambda')-\int\frac{d\mu}{2\pi}\varphi(\lambda-\mu)n(\mu)[\varphi^\text{dr}]_{\mu}^{\, \lambda'}.
\ee

Whereas Eulerian GHD is most easily solved as infinitesimal translations of the filling function, a stable strategy to solve convection-diffusion equations is the Crank-Nicholson method \cite{burmeister1993convective}, which averages between the forward and backward Euler method. Let us compactly write the GHD equations as $\partial_t \rho= F[\rho]$, with $F$ grouping together all the terms in \eqref{eq_diff_root}.
Then we update the time evolution of the root density as
\be
\rho_{t+\Delta t}= \rho_{t}+\Delta t\, F\left[\frac{\rho_{t+\Delta t}+\rho_{t}}{2}\right]\, .
\ee
This method is implicit and the above equation is iteratively solved at each time step, until convergence is attained, as it has been done in Ref. \cite{PhysRevLett.125.240604} to study diffusion in a trapped Lieb-Liniger model.
The presence of diffusion, besides being a physical effect captured by the GHD equations, takes an important role in making the numerical method stable: diffusive terms smooth out the root density profile, preventing it from developing singularities as in the pure convective case \ref{sec_firstorder_lackth}. If the diffusion term is too small compared with the convective one, the roughness of the root density grows with time eventually causing instabilities in the derivative discretization, similarly to what happens to the limit of large Reynolds numbers in standard hydrodynamics.

\bibliography{biblio}

\end{document}